\documentclass[11pt]{article}
\usepackage{graphicx}
\usepackage{amssymb,amsmath}
\usepackage{xcolor}
\usepackage[sort&compress,numbers]{natbib}
\usepackage{bm}

\begin{document}

\title{
\textbf{Corrections to scaling in the 2D $\varphi^4$ model:
Monte Carlo results and some related problems}
}

\author{J. Kaupu\v{z}s$^{1,2}$ 
\thanks{E--mail: \texttt{kaupuzs@latnet.lv}}
\hspace{1ex},
R. V. N. Melnik$^{2,3}$ \\
$^1$ Institute of Technical Physics, Faculty of Natural Sciences and Technology, \\
Riga Technical University,  P. Valdena 3/7, LV-1048 Riga, Latvia \\
$^2$ The MS2 Discovery Interdisciplinary Research Institute, \\
Wilfrid Laurier University, Waterloo, Canada
\\
$^3$ BCAM - Basque Center for Applied Mathematics, E48009 Bilbao, Spain}

\maketitle

\begin{abstract}
Monte Carlo (MC) simulations have been performed to refine the estimation of the correction-to-scaling exponent $\omega$ in the 2D $\varphi^4$ model, which belongs to one of the most fundamental universality classes.
If corrections have the form $\propto L^{-\omega}$,
then we find $\omega=1.546(30)$
and $\omega=1.509(14)$ as the best estimates.
These are obtained from the finite-size scaling of
the susceptibility data in the range of linear lattice sizes $L \in [128,2048]$
at the critical value of the Binder cumulant
and from the scaling of the corresponding pseudocritical couplings
within $L \in [64,2048]$.
These values agree with several other MC estimates at the assumption of the
power-law corrections and are
comparable with the known results of the $\epsilon$-expansion.
In addition, we have tested the consistency with the scaling corrections
of the form $\propto L^{-4/3}$, $\propto L^{-4/3} \ln L$ and
$\propto L^{-4/3} /\ln L$, which might be expected from some
considerations of the renormalization group and Coulomb gas model.
The latter option is consistent with our MC data.
Our MC results served as a basis
for a critical reconsideration of some earlier theoretical conjectures and
scaling assumptions. In particular, we have corrected and refined our previous analysis by grouping Feynman diagrams. The renewed analysis gives
$\omega \approx 4-d-2 \eta$ as some approximation
for spatial dimensions $d<4$, or $\omega \approx 1.5$ in two dimensions.
\end{abstract}

\hspace*{8ex}
 renormalization group, Coulomb gas, Feynman diagrams, $\epsilon$-expansion

\section{Introduction}
\label{intro}

The critical phenomena and critical exponents have been extensively investigated during
many decades, using the $\varphi^4$ (or Ginzburg-Landau) model as one of the basic models
for analytical~\cite{Amit,Ma,Justin,Kleinert,PV,K_Ann01,Shalaby}, as well as
numerical~\cite{MHB86,TC90,MF92,our2D,AGKKW17,AKY21,nacom24} studies.
We consider the local order parameter $\varphi$ as a scalar or, more generally,
an $N$-component vector. In the latter case, this $\varphi^4$ model is known also as the
$O(N)$ model, pointing to its rotational symmetry.
It represents the most celebrated example of universality~\cite{nacom24}, which describes
the critical properties of a wide range of physical phase transitions,
such as ferromagnetic and anti-ferromagnetic,
liquid-vapour, as well as superconducting and superfluid transitions.

The scalar $\varphi^4$
model belongs to the Ising universality class. The Ising model is more convenient for numerical
studies than the $\varphi^4$ model, therefore most of the numerical results for this
universality class come just from the Ising model. The critical exponents of the 2D Ising model are known exactly~\cite{Onsager,Baxter}. Hence, the numerical studies
have been mainly focused on the 3D case, including Monte Carlo (MC)
simulations~\cite{HasRev,Ferrenberg,mansMC3D}, high temperature (HT) series
expansion~\cite{Wipf,BC2001,HTCompostrini} and the conformal bootstrap
calculations~\cite{Showk,bootstrap,CFTrecent,CFK22}.

Although the 2D Ising model is well studied, we have found that the
2D case of the scalar $\varphi^4$ model is very intriguing, which requires a further investigation. A point here is that the 2D $\varphi^4$ model might contain
non-trivial corrections to scaling, which do not show up or vanish in the 2D Ising model, as regards a subset of quantities, including the moments of magnetization
and energy per spin, as well as other related quantities.
This possibility has been pointed out in our earlier study~\cite{our2D},
based on MC simulations and
some analytical arguments. Historically, several results have been obtained for the
correction-to-scaling exponent $\omega$ in the 2D $\varphi^4$ model, which differ
from $\omega=2$ usually observed in the 2D Ising model. In particular,
$\omega=4/3$ has been conjectured in~\cite{Justin}, based on the perturbative
renormalization group (RG) calculations. Later, this discrepancy with
$\omega=2$
has been explained in~\cite{PVx}, suggesting that the Callan-Symanzik $\beta$-function,
used in these RG calculations, is singular at the fixed point.
However, recent results of the
$\epsilon$-expansion, which do not use this $\beta$-function, also give somewhat
smaller than $2$ value of $\omega$, i.~e., $\omega=1.71(10)$~\cite{Shalaby}.

Non-integer correction-to-scaling exponents appear in the 2D Ising model,
considering a wider set of quantities. In particular, the fractal geometry of
the critical Potts clusters has been studied in~\cite{AA} within the framework
of RG and Coulomb gas approach, considering the
fractal dimensions $D_M$, $D_H$, $D_{EP}$ and $D_{SC}$, describing the scaling of
cluster's mass, hull, externally accessible perimeter and singly connected bonds, respectively,
with its radius of gyration $R$. It has been found that the scaling corrections for the effective fractal dimensions
are described by certain exponents $\theta$, $\theta'$,
$\theta''$ and 1.
As mentioned in~\cite{AA}, it is reasonable to interpret $\theta'$
and $\theta''$
as the scaling exponents of some irrelevant perturbations in the spirit of the renormalization group.
Hence, these correction exponents can affect also other quantities. The exact values
of $\theta$ and $\theta'$ in the $q$-state Potts model at $q=2$,
i.~e., in the Ising model, are $\theta=\theta'=4/3$~\cite{AA}. In view of these arguments,
the correction-to-scaling exponent $4/3$ can generally exist in
the models of the 2D Ising universality class, including the
2D $\varphi^4$ model. Moreover, the coincidence of the two exponents $\theta$ and $\theta'$
is a special case,
therefore one may look for possible logarithmic corrections.

In the current work, we have performed
much more accurate (longer) MC simulations than those in~\cite{our2D} to
estimate $\omega$ in the case of pure power-law corrections to scaling,
as well as to test the influence of logarithmic corrections
and consistency with the exponent $\omega=4/3$ with and without the logarithmic corrections.
As a result, we have found $\omega=1.509(14)$ as our best estimate at the assumption
of pure power-law corrections. This value is shifted to $\omega \approx 7/4$,
assuming the logarithmic correction of the form $\ln L$, and to
$\omega \approx 4/3$, assuming the $1/\ln L$ correction. In any way, these values appear to be  somewhat
surprising from the point of view of our earlier scaling arguments and analysis~\cite{our2D}, where the existence of a correction-to-scaling with
exponent $3/4$ was expected. Hence, it served as a basis for a
reconsideration of our earlier scaling assumptions. More flexible scaling
assumptions allowed us to resolve this issue, as well as to correct and refine
the analysis by grouping of Feynman diagrams initiated in~\cite{K_Ann01}.

The paper is organized as follows. The results of MC simulation
are reported in Sec.~\ref{sec:MCs}. The estimation of $\omega$
from our MC data and the comparison with some of the known results
are provided in Sec.~\ref{sec:MCes}.
A critical reconsideration of the scaling arguments
in~\cite{our2D} is provided in
Sec.~\ref{sec:CV}, and the related renewed analysis by grouping of
Feynman diagrams is considered in Sec.~\ref{sec:GFD}.
Finally, the summary and discussion is given in Sec.~\ref{sec:summary}

\section{Monte Carlo simulation results for the lattice $\varphi^4$ model}
\label{sec:MCs}

We have performed MC simulations of the 2D $\varphi^4$ model on square lattice
with periodic boundary conditions. The Hamiltonian $\mathcal{H}$ is given by
\label{sec:model}
\begin{equation}
\frac{\mathcal{H}}{k_{\mathrm{B}} T}  = - \beta \sum\limits_{\langle i j \rangle}
\varphi_i \varphi_j + \sum\limits_i \left( \varphi_i^2 + \lambda \left( \varphi_i^2 -1 \right)^2 \right) \;,
\end{equation}
where $T$ is the temperature $k_{\mathrm{B}}$ is the Boltzmann constant,
$-\infty < \varphi_i < \infty$ is a continuous scalar order parameter at the $i$-th lattice site, and $\langle ij \rangle$
denotes the set of all nearest neighbors. Here we use the same notations as in our
earlier MC study of this model~\cite{our2D}, related to those in~\cite{Hasenbusch} via
$\beta= 2 \kappa$ and $\varphi = \phi$.
The simulations have been performed by the hybrid algorithm, where cluster updates are
combined with Metropolis sweeps, as described in~\cite{Hasenbusch,our2D}.
In our simulations, one step of the hybrid algorithm consisted of
one Metropolis sweep and $N_W$ clusters of the Wolff algorithm. The parameter
$N_W$ was chosen in such a way to ensure that
the entire number of spin moves in $N_W$ clusters is
about $2/3$ or $0.6$ of the total number of spins.

The simulations in~\cite{our2D} have been performed at $\lambda = 0.1$,
$\lambda = 1$ and $\lambda = 10$
at pseudocritical couplings $\widetilde{\beta}_c(L)$, corresponding to
$U=\langle m^4 \rangle / \langle m^2 \rangle^2 = 1.1679229 \approx U^*$
and $U=2$, where $m$ is the magnetization per spin, $1-U/3$ is the Binder cumulant,
and $U^*$ is the known critical value of $U$ found in~\cite{Sokal}.
Here we have performed extended simulations at $\lambda = 0.1$, making them
significantly longer and extending the range of linear lattice sizes from $L \le 1536$
to $L \le 2048$. In~\cite{our2D}, the results for each lattice size have been obtained
from $100$ simulation bins (iterations), each consisting of $10^6$ hybrid algorithm steps.
In the extended here simulations, the latter number is increased to $3\cdot 10^6$ steps,
the number of simulation bins being chosen $300$ for $L\in [6,384]$, $200$ for $L=512$,
$100$ for $L \in [768,1536]$ and $120$ for $L=2048$.
The final results for the overlapping lattice sizes within $L \le 1536$ have been obtained
by a weighted averaging of the actual MC values and those in~\cite{our2D}, applying the
weight coefficients proportional to the simulation length (the number of MC steps).
The results for the normalized susceptibility $\chi/L^{7/4}$
(where $\chi = L^2 \langle m^2 \rangle$) and the normalized derivative
$-(\partial U /\partial \beta)/L$ are collected in Tab.~\ref{tab1} and Tab.~\ref{tab2}
at pseudocritical couplings $\widetilde \beta_c(L)$, corresponding to
$U=U^*$  and $U=2$,
respectively. The used here normalization ensures that these quantities tend to certain
finite values at $L \to \infty$, in accordance with the known critical exponents
$\gamma=7/4$ and $\nu=1$ of the 2D Ising universality class.

\begin{table}
\caption{The values of $\widetilde \beta_c(L)$, $\chi/L^{7/4}$ and
$-(\partial U /\partial \beta)/L$ for $\lambda=0.1$ and $U=1.1679229 \approx U^*$
depending on the lattice size $L$.}
\label{tab1}
\begin{center}
\begin{tabular}{|c|c|c|c|}
\hline
\rule[-2mm]{0mm}{7mm}
$L$ & $\widetilde \beta_c$ & $\chi/L^{7/4}$  & $-(\partial U /\partial \beta)/L$  \\
\hline
6 & 0.6310451(76) & 1.92054(13) & 0.86991(19)  \\
8 & 0.6205904(55) & 1.70513(11) & 0.91937(19)  \\
12 & 0.6126206(35) & 1.494396(97) & 0.98891(23) \\
16 & 0.6097694(27) & 1.394495(83) & 1.03316(23) \\
24 & 0.6077895(17) & 1.302878(78) & 1.08511(27) \\
32 & 0.6071420(14) & 1.261811(75) & 1.11200(29) \\
48 & 0.60673002(87) & 1.226245(65) & 1.14005(32)  \\
64 & 0.60660154(68) & 1.210966(61) & 1.15320(34)  \\
96 & 0.60652379(44) & 1.198190(59) & 1.16602(37) \\
128 & 0.60650059(31) & 1.192741(53) & 1.17128(36) \\
192 & 0.60648656(21) & 1.188314(50) & 1.17676(38) \\
256 & 0.60648299(17) & 1.186707(47) & 1.17978(37) \\
384 & 0.60648037(11) & 1.185094(49) & 1.18112(38) \\
512 & 0.606479833(90) & 1.184499(54) & 1.18115(45) \\
768 & 0.606479199(84) & 1.183966(72) & 1.18249(67) \\
1024 & 0.606479261(66) & 1.183828(71) & 1.18258(72) \\
1536 & 0.606479163(42) & 1.183637(62) & 1.18270(72) \\
2048 & 0.606479052(31) & 1.183539(72) & 1.18202(76) \\
\hline
\end{tabular}
\end{center}
\end{table}

\begin{table}
\caption{The same quantities as in Tab.~\ref{tab1} for $\lambda=0.1$ and $U=2$.}
\label{tab2}
\begin{center}
\begin{tabular}{|c|c|c|c|}
\hline
\rule[-2mm]{0mm}{7mm}
$L$ & $\widetilde \beta_c$ & $\chi/L^{7/4}$  & $-(\partial U /\partial \beta)/L$  \\
\hline
6 & 0.5622957(89) & 0.500854(71) & 2.50489(62)  \\
8 & 0.5705578(64) & 0.446877(65) & 2.54550(65)  \\
12  & 0.5804632(43) & 0.394622(60) & 2.59107(77)  \\
16  & 0.5861528(33) & 0.370388(59) & 2.61472(77)  \\
24  & 0.5924095(22) & 0.349474(55) & 2.64609(86)  \\
32  & 0.5957579(16) & 0.341174(51) & 2.66479(96)  \\
48  & 0.5992428(11) & 0.335385(49) & 2.68905(95)  \\
64  & 0.60102839(66) & 0.333833(43) & 2.70895(94)  \\
96  & 0.60283621(49) & 0.333458(43) & 2.7284(10)  \\
128  & 0.60374599(36) & 0.333854(45) & 2.7382(11)  \\
192  & 0.60465782(23) & 0.334880(38) & 2.7563(11)  \\
256  & 0.60511400(17) & 0.335550(37) & 2.7625(11)  \\
384  & 0.60556994(12) & 0.336520(36) & 2.7729(11) \\
512  & 0.60579760(11) & 0.336986(45) & 2.7772(15) \\
768  & 0.606024909(96) & 0.337482(55) & 2.7827(20) \\
1024   & 0.606138639(68) & 0.337825(56) & 2.7831(19) \\
1536   & 0.606252317(45) & 0.338269(48) & 2.7885(18) \\
2048   & 0.606308994(35) & 0.338425(52) & 2.7904(21) \\
\hline
\end{tabular}
\end{center}
\end{table}

\section{MC estimation of the correction-to-scaling exponent $\omega$}
\label{sec:MCes}

\subsection{Estimations from the susceptibility data}
\label{sec:chi}

According to the finite-size scaling theory, the susceptibility behaves as
$\chi \propto L^{\gamma/\nu} f(L/\xi)$ for large $L$ in vicinity of the critical point,
where $\xi$ is the correlation length and $f(L/\xi)$ is the scaling function.
This is the leading asymptotic behaviour, but
we are particularly interested in corrections to scaling. There are generally
(at arbitrary spatial dimension $d$) non-analytical corrections to scaling,
which are by a factor $\mathcal{O}\left(L^{-\omega} \right)$ smaller than
the leading term, $\omega$ being the leading correction-to-scaling exponent,
as well as multiplicative and additive analytical corrections.
The multiplicative analytical correction can be represented by the correction factor
$1 + \mathcal{O}(t)$, where $t=(\beta_c-\beta)/\beta_c$ is the reduced temperature
and $\beta_c$ is the critical coupling, whereas the additive analytical correction
represents an analytical background contribution. Noting that
$\widetilde{\beta}_c(L)-\beta_c \sim L^{-1/\nu}$
holds for the pseudocritical coupling at $U=2$,
we have $\mathcal{O}(t) \sim L^{-1/\nu}$ in this scaling regime.
The case $U=U^*$ is special, where the amplitude at $L^{-1/\nu}$ vanishes.
These scaling considerations are consistent with the fact that
$U= F(t L^{1/\nu})$ holds asymptotically with some regular
scaling function $F$~\cite{HasRev}, and $U^*=F(0)$. In the first approximation,
the additive analytical background contribution can be approximated by a constant
value $\chi_0$, determined at $t=0$ in the thermodynamic limit.
Hence, neglecting corrections of
higher orders, the normalized susceptibility of the 2D $\varphi^4$ model
at $\beta=\widetilde{\beta}_c(L)$ can be approximated as
\begin{equation}
 \chi/L^{7/4} = A + B L^{-\omega} + C L^{-1} + \chi_0 L^{-7/4} \;,
\label{eq:chifit0}
 \end{equation}
noting that $C=0$ holds at $U=U^{*}$.
This ansatz assumes pure power-law corrections and should be modified as
\begin{equation}
 \chi/L^{7/4} = A + B L^{-\omega} \, (\ln L)^{\kappa}+ C L^{-1} + \chi_0 L^{-7/4} \,,
\label{eq:chifit0log}
 \end{equation}
with $\kappa=0,\pm1$ to include possible logarithmic
corrections discussed in Sec.~\ref{intro}.

We start our analysis with the test of pure power-law scenario ($\kappa=0$),
setting $\chi_0$ to zero at the beginning.
The validity of such fits is confirmed a posteriori, as they provide stable results
with $\omega<7/4$, implying that $\chi_0 L^{-7/4}$ is a smaller term than $B L^{-\omega}$
at $L \to \infty$. In the case of $U=U^*$, it reduces to three-parameter fits
to the ansatz $\chi/L^{7/4} = A + B L^{-\omega}$. The results for $\omega$, obtained
by such fitting over the range of sizes
$L \in [L_{\mathrm{min}},2048]$, are shown in Tab.~\ref{tab3} depending on
$L_{\mathrm{min}}$. The quality of fits is controlled by the value of
$\chi^2/\mathrm{d.o.f.}$ ($\chi^2$ of the fit per degree of freedom), as well as by
the goodness $Q$~\cite{Recipes}. A smaller $\chi^2/\mathrm{d.o.f.}$
and a larger $Q$ (where $0 < Q < 1$) implies a higher quality of the fit,
noting that $\chi^2/\mathrm{d.o.f.}$ is about unity for moderately good fits,
and fits with $Q>0.1$ are usually considered as acceptable.
According to these criteria, the fits with $L_{\mathrm{min}}$ from $64$ to $128$
in Tab.~\ref{tab3} are acceptable. The corresponding values of $\omega$
agree within the error bars. We assume the estimate $\omega = 1.546(24)$
at $L_{\mathrm{min}}=128$
as the best one, as it has a significantly smaller $\chi^2/\mathrm{d.o.f.}$ and
larger $Q$ values than in other cases, noting that further increasing of
$L_{\mathrm{min}}$ only increases the statistical errors bars without improvement
of the fit quality.

\begin{table}
\caption{The MC estimates of $\omega$, obtained by fitting the susceptibility
data in Tab.~\ref{tab1} to the ansatz $\chi/L^{7/4} = A + B L^{-\omega}$ within
$L \in [L_{\mathrm{min}},2048]$. The quality of theses fits is controlled
by  the $\chi^2/\mathrm{d.o.f.}$ and the goodness $Q$ values in the last two columns.}
\label{tab3}
\begin{center}
\begin{tabular}{|c|c|c|c|}
\hline
\rule[-2mm]{0mm}{7mm}
$L_{\mathrm{min}}$ & $\omega$ & $\chi^2/\mathrm{d.o.f.}$ & $Q$ \\
\hline
32  &  1.5124(27)  &  7.92  &  $7 \cdot 10^{-12}$ \\
48  &  1.5454(50)  &  1.85  &  0.055 \\
64  &  1.5567(78)  &  1.61  &  0.115  \\
96  &  1.577(16)   &  1.52  &  0.154  \\
128 &  1.546(24)   &  1.34  &  0.237  \\
\hline
\end{tabular}
\end{center}
\end{table}

\begin{table}
\caption{The MC estimates of $\omega$, obtained by fitting the susceptibility
data in Tab.~\ref{tab2} to the ansatz $\chi/L^{7/4} = A + B L^{-\omega} + C L^{-1}$ within
$L \in [L_{\mathrm{min}},2048]$. The quality of theses fits is controlled
by  the $\chi^2/\mathrm{d.o.f.}$ and the goodness $Q$ values in the last two columns.}
\label{tab4}
\begin{center}
\begin{tabular}{|c|c|c|c|}
\hline
\rule[-2mm]{0mm}{7mm}
$L_{\mathrm{min}}$ & $\omega$ & $\chi^2/\mathrm{d.o.f.}$ & $Q$ \\
\hline
24  &  1.269(14)  &  3.98 &  $1.9 \cdot 10^{-5}$ \\
32  &  1.333(20)  &  2.37 &  0.011 \\
48  &  1.431(40)  &  1.61 &  0.117 \\
64  &  1.540(69)  &  1.24 & 0.276  \\
\hline
\end{tabular}
\end{center}
\end{table}

In the case of $U=2$, the ansatz $\chi/L^{7/4} = A + B L^{-\omega}$ does not
provide acceptable fits because $\chi/L^{7/4}$ is a nonmonotoneous
function of $L$, as it can be seen from Tab.~\ref{tab2}. Inclusion of the expected
in this case analytical correction term $C L^{-1}$ solves the problem. The corresponding
fit results are collected in Tab.~\ref{tab4}. By the same analysis as before,
we find $\omega=1.540(69)$ as the best estimate in this case. The optimal
minimal lattice size
$L_{\mathrm{min}}=64$ appears to be smaller than $128$ found for $U=U^*$.
It is explained by the fact
that the number of fit parameters is increased and, therefore, only fits
over a wider range of sizes appear to be reasonably stable.

Our best MC estimates at $U=U^*$ and $U=2$  are well consistent. However,
there might be some concerns about the possible systematic errors due to
the neglection of the constant background contribution, since the exponent $-7/4$
in~(\ref{eq:chifit0}) is quite close to our actual estimates of $-\omega$.
Unfortunately, the inclusion of $\chi_0$ as an extra fit parameter does not lead
to accurate and stable results because of too many fit parameters.
Nevertheless, the possible effect of the background term can be tested
by considering fits with several fixed $\chi_0$ values within some range
of reasonable values. To get an idea about the possible values of $\chi_0$,
we refer to the known results for the 2D Ising model~\cite{Perk}, i.e.,
$\chi_0 \approx -0.1041$,  $\chi_0 \approx -0.0496$
and $\chi_0 \approx -0.2215$ for the ferromagnetic case ($\beta>0$) on square, triangular
and honeycomb lattices, respectively, as well as
$\chi_0 \approx 0.1589$ and $\chi_0 \approx 0.1224$ for the antiferromagnetic case
($\beta<0$) on square and honeycomb lattices.
Recall that the Ising model corresponds to the large-$\lambda$ limit of the actual
$\varphi^4$ model. Hence, the $\chi_0$ value for $\lambda=10$
could be quite close
to the Ising value $-0.1041$ on the actually considered square lattice.
Since the susceptibility
$\chi$ does not change much with $\lambda$ (as it can be seen from the plots
for $\lambda = 0.1, 1$ and $10$ in~\cite{our2D}) and $\chi_0$ in different cases
(referred above) also are quite similar in magnitude,
we do not expect a much larger
$\mid \chi_0 \mid$ value in the actual case of $\lambda=0.1$.
The exponent $\omega$ is changed only very slightly, i.~e.,
from $\omega=1.5458(244)$ to $\omega=1.5463(244)$, if the fit with $L_{\mathrm{min}}=128$
at $U=U^*$ is biased by adding the background term with $\chi_0=-0.1$.
The influence on the best fit at $U=2$ is slightly larger, showing up as a shift by $0.0011$.
Thus, the fit at $U=U^*$ provides the most accurate and stable value
obtained from the susceptibility data if corrections
to scaling have the power-law form. We have rounded up
its error bars to $0.030$ for our final best estimation from these
data, i.~e.,
\begin{equation}
 \omega = 1.546(30)
 \label{eq:omega}
\end{equation}
to include the statistical error of one
standard deviation ($\sigma$),
as well as the additional uncertainty due to the $\chi_0 L^{-7/4}$ term
in~(\ref{eq:chifit0}) if $\mid \chi_0 \mid <1$.
According to our discussion, it is plausible that $\mid \chi_0 \mid$
is not larger than $1$.
However, due to
the lack of precise information about $\chi_0$, the error bars in~(\ref{eq:omega})
are indicative.

If the restriction $\mid \chi_0 \mid <1$ is removed, the uncertainty in $\omega$
becomes significantly larger. In particular, the fit to
$\chi/L^{7/4} = A + BL^{-\omega} + \chi_0 L^{-7/4}$
over $L \in [96,2048]$ with fixed $\omega=4/3$  is relatively good, as we have
$\chi^2/\mathrm{d.o.f}=1.32$ and $Q=0.236$ in this case.
 It provides some evidence that $\omega$ could be consistent with $4/3$ reported in~\cite{AA}. However, an acceptable fit is possible also at a different
 $\omega$ value, e.~g., at $\omega=1.5$. At $\omega=4/3$,
the fitted value of $\chi_0$, i.~e., $\chi_0= 26.4(1.6)$, greatly exceeds in magnitude
the corresponding value $\chi_0 \approx -0.1041$
of the 2D Ising model. Moreover, in the 2D Ising case, $\chi_0$ appears to be always
(for all lattices considered in~\cite{Perk}) negative for the ferromagnetic interaction and
positive for the antiferromagnetic one. Since we have the ferromagnetic interaction,
the positive value of $\chi_0= 26.4(1.6)$ does not fit in this qualitative picture. Hence,
the validity of the fit with $\omega=4/3$ seems doubtful from an intuitive point of view, at least.
Another aspect is that the estimate~(\ref{eq:omega}) agrees well with
several other estimates, summarized in Sec.~\ref{sec:MCcom}, obtained under the assumption of a pure power-law scaling
of the correction terms. This agreement with~(\ref{eq:omega})
and disagreement with $\omega=4/3$ of all these estimates could not
be accidental.
Hence, if we have the pure power-law scaling form, then $\omega$ is about $1.546$ rather than $4/3$.
Finally, although the estimated error bars are never rigorous due to the possible influence
of the correction terms not included in the fit ansatz, the good consistency of the
results serves as a reasonable evidence that these error bars are not essentially underestimated.

The estimated value of $\omega$ changes if we include logarithmic corrections.
In particular, our best estimate $\omega=1.546(24)$ in Tab.~\ref{tab3} changes
to $1.348(25)$ (with $\chi^2/\mathrm{d.o.f.}=1.32$ and $Q=0.243$) for the $1/\ln L$
logarithmic correction ($\kappa=-1$), and to $1.744(24)$
(with $\chi^2/\mathrm{d.o.f.}=1.38$ and $Q=0.218$) for the $\ln L$
logarithmic correction ($\kappa=1$). Thus, the data appear to be consistent with
$\omega=4/3$ in the first case and with $\omega=7/4$ in the second case.
We note that the first value might be expected according to the results
of~\cite{AA} discussed in Sec.~\ref{intro}, whereas the second value
agrees well with the result $\omega=1.71(10)$ of the $\epsilon$-expansion
in~\cite{Shalaby}, where $1.75$ has been considered as the exact value.

\begin{figure}
\begin{center}
\includegraphics[width=0.9\textwidth]{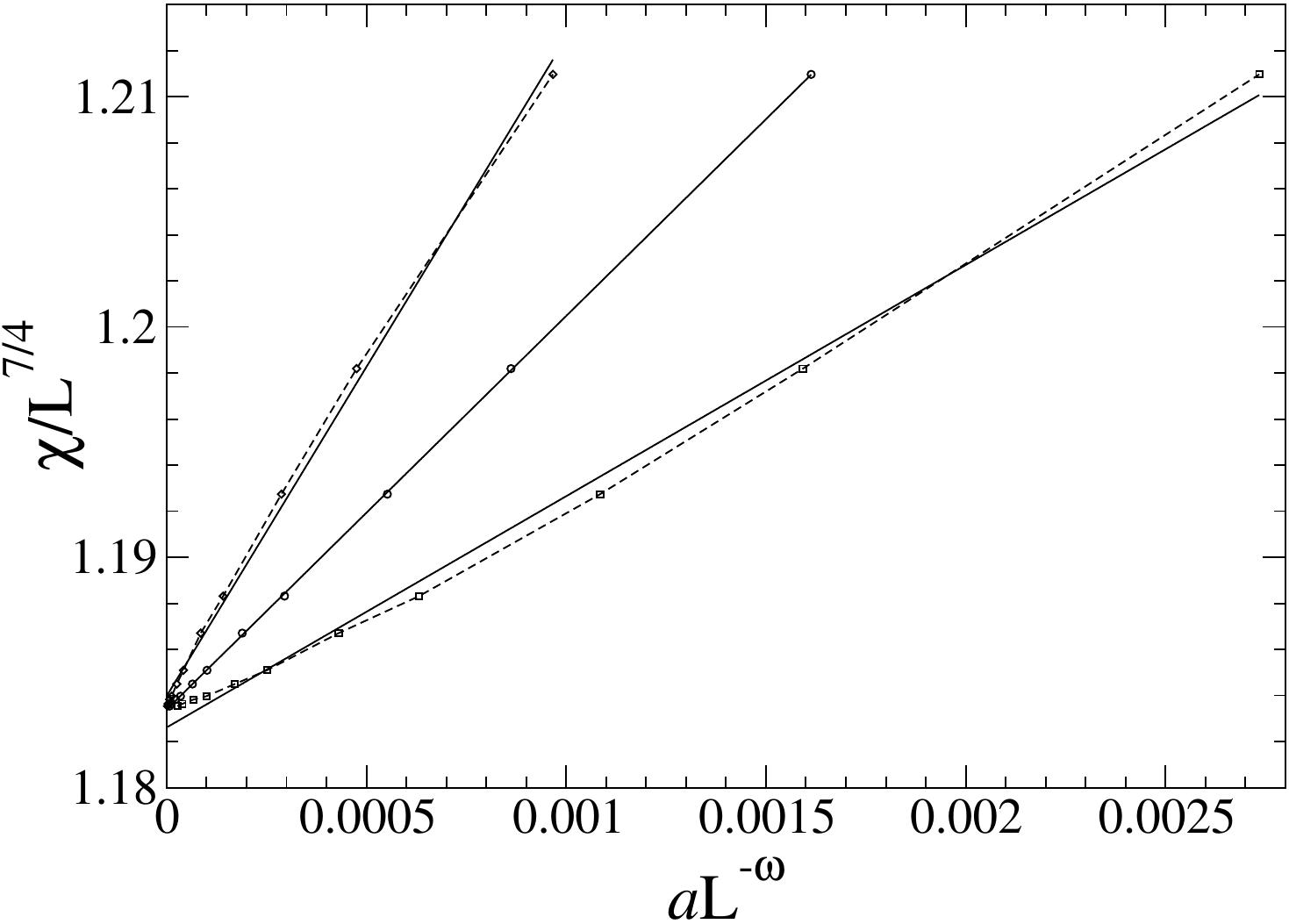}
\end{center}
\caption{The $\chi/L^{7/4}$ vs $aL^{-\omega}$ plots
within $L \in [64,2048]$ for $U=U^*$ at
$a=0.7$, $\omega=4/3$ (squares); $a=1$,
$\omega=1.546$ (circles); and $a=1.4$, $\omega=7/4$
(diamonds). The statistical errors of one $\sigma$
are smaller than the symbol size.
The linear fits are shown by straight lines.}
\label{fig1}
\end{figure}

Despite of this variation in $\omega$ depending on $\kappa$
in the fit ansatz~(\ref{eq:chifit0log}), the uncertainty in $\omega$ is
quite small for our estimation procedure at a fixed $\kappa=0$ and $U=U^*$,
assuming a small influence of the $\chi_0 L^{-7/4}$ term
in agreement with the arguments provided before.
It is manifested in~(\ref{eq:omega}) and can be concluded also from the
$\chi/L^{7/4}$ vs $aL^{-\omega}$ plots (where $a$ is a
scaling factor, different for each $\omega$)
at $\omega=4/3, 1.546$ and $7/4$ shown in Fig.~\ref{fig1} for the range of sizes
$L \in [64,2048]$.
The $\chi/L^{7/4}$ vs $aL^{-\omega}$
plot is almost perfectly linear at $\omega=1.546$ and is essentially curved at $\omega=4/3$.
It indicates that the uncertainty in the estimation of $\omega$ at
fixed $\kappa=0$
is essentially smaller than the difference between $1.546$ and $4/3$.

\begin{figure}
\begin{center}
\includegraphics[width=0.9\textwidth]{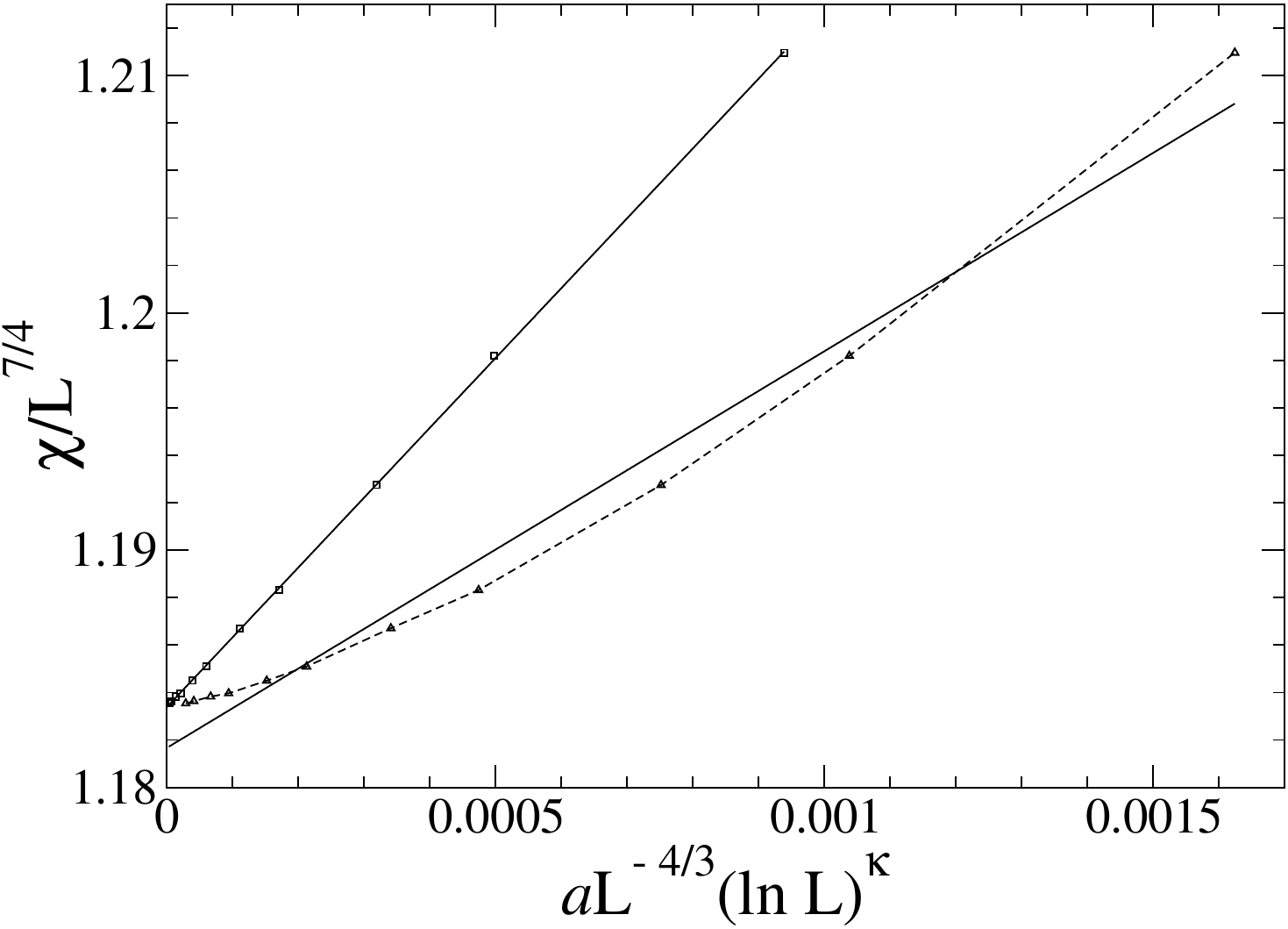}
\end{center}
\caption{The $\chi/L^{7/4}$ vs $aL^{-4/3} (\ln L)^{\kappa}$ plots within $L \in [64,2048]$ for $U=U^*$ at
$a=1$, $\kappa=-1$ (squares) and  $a=0.1$,
$\kappa=1$ (triangles). The statistical errors of one $\sigma$
are smaller than the symbol size.
The linear fits are shown by straight lines.}
\label{fig2}
\end{figure}

Since the logarithmic corrections can be expected at $\omega=4/3$,
we have presented the plots for $\chi/L^{7/4}$ vs
$aL^{-4/3} (\ln L)^{\kappa}$ with $\kappa = \pm 1$ in Fig.~\ref{fig2}.
It is evident from this figure that the
$\propto L^{-4/3} \ln L$ scaling ($\kappa=1$) is even
worse consistent with the data than the $\propto L^{-4/3}$ scaling
tested in Fig.~\ref{fig1}. To the contrary, the plot with $\kappa=-1$ in Fig.~\ref{fig2}
looks almost perfectly linear. The $\chi^2/{\mathrm{d.o.f.}}$ values of the corresponding linear fits within $L \in [L_{\mathrm{min}},2048]$
are $1.89$, $1.82$ and $1.18$ for $L_{\mathrm{min}}=64, 96$ and $128$,
respectively. The corresponding $Q$ values are $0.049$, $0.069$ and
$0.31$. It means that the fit within $L \in [128,2048]$ is well acceptable
(even slightly better than the fit with the $L^{-\omega}$ correction)
and the $\propto L^{-4/3}/\ln L$
correction-to-scaling behavior is
consistent with our data for $L \ge 128$ at $U=U^*$. The consistency
at $U=2$ has also been verified. The fit
of these data to the ansatz
$\chi/L^{7/4} = A + B L^{-4/3}\,(\ln L)^{-1} + CL^{-1}$
within $L \in [64,2048]$ with $\chi^2/{\mathrm{d.o.f.}}=1.30$
and $Q=0.24$ is almost as good as the corresponding
fit to the ansatz $\chi/L^{7/4} = A + B L^{-\omega} + CL^{-1}$
(see Tab.~\ref{tab4}). One can also fit the same data set
to $\chi/L^{7/4} = A + B L^{-7/4}\,\ln L + CL^{-1}$
with $\chi^2/{\mathrm{d.o.f.}}=1.33$ and $Q=0.225$.

\subsection{Estimation from the $\partial U/\partial \beta$ data}
\label{sec:Uat}

Similar scaling arguments as in Sec.~\ref{sec:chi} can be applied also to
the derivative $\partial U/\partial \beta$ with the only difference that the additive
constant analytical background contribution is not expected in this case,
since  $\partial U/\partial \beta$
is strictly zero in the thermodynamic limit at $\beta \ne \beta_c$.
It is confirmed by the fits of our $-(\partial U/\partial \beta)/L$ data for $U=U^*$.
The additive background term would give a $\sim L^{-1}$ correction,
which is not observed in this case. Such a correction appears at $U\ne U^*$
as a multiplicative analytical correction. Hence, neglecting corrections of higher
orders, $-(\partial U/\partial \beta)/L$ at $\beta = \widetilde{\beta}_c(L)$ can be
approximated by
\begin{equation}
 -(\partial U/\partial \beta)/L = A + B L^{-\omega} + C L^{-1} \,,
 \label{eq:Uat}
\end{equation}
with $C=0$ for $U=U^*$. In the case of the logarithmic corrections,
this ansatz has to be modified as
\begin{equation}
 -(\partial U/\partial \beta)/L = A + B L^{-\omega} \,
 (\ln L)^{\kappa} + C L^{-1} \;.
 \label{eq:Uatlog}
\end{equation}

First, we test the pure power-law scaling using~(\ref{eq:Uat}).
Thus, for  $U=U^*$, a simple ansatz
$-(\partial U/\partial \beta)/L = A + B L^{-\omega}$ is applicable
in this case, and the corresponding
fit results are shown in Tab.~\ref{tab5}. The obtained values of $\omega$, however, are
varied remarkably with the minimal lattice size $L_{\mathrm{min}}$ used in the fits.
The results are stabilized, including the expected next correction
$\propto L^{-2}$. Such fits have been already considered
in~\cite{our2D}, providing $\omega = 1.373(48)$ by fitting over the range
$L \in [12,1536]$ with $\chi^2/\mathrm{d.o.f} = 1.5$ and $Q=0.123$. The poor and marginal
quality of those fits in~\cite{our2D} raised doubts about the validity of the
ansatz used. This situation is improved essentially by the actual refined data
in Tab.~\ref{tab1}. The fit at $L_{\mathrm{min}}=12$ is poor and gives a similar $\omega$
value than that one reported in~\cite{our2D}. However,
we obtain moderately good and consistent fits
for $L_{\mathrm{min}}=24$ and $L_{\mathrm{min}}=32$, as
shown in Tab.~\ref{tab6}. Hence, we report $\omega=1.567(58)$ as the best estimate
from the $\partial U/\partial \beta$ data with $U=U^*$, if
the non-analytical corrections to scaling have the pure
power-law form, because the results of this
fitting procedure are reasonably stable and the corresponding fit in Tab.~\ref{tab6}
at $L_{\mathrm{min}}=24$ is well acceptable and has the highest quality according to
the $\chi^2/\mathrm{d.o.f}$ and $Q$ criteria.

We have performed fits of the $-(\partial U/\partial \beta)/L$ data for
$U=2$ in Tab.~\ref{tab2}, starting with a simple ansatz
$-(\partial U/\partial \beta)/L = A + B L^{-\omega}$ to compare the results with
those in~\cite{our2D}. Such fits over $L \in [16,1536]$
in~\cite{our2D} provided an evidence that $\omega$, probably, is $1/2$.
Having extended and more accurate actual data, we have tested how results
change for significantly larger than $16$ values of $L_{\mathrm{min}}$, as shown in
Tab.~\ref{tab7}. It can be seen that $\omega$ about $1/2$ describes a transient
behavior, since the estimated value of $\omega$ tends to increase with
$L_{\mathrm{min}}$, perhaps, to the value $\omega=1$, provided by the analytical
correction term $\sim L^{-1}$. In this case, one needs to include this correction
term into the ansatz to extract the non-analytical correction exponent, if it
exists. Hence, we have used the ansatz~(\ref{eq:Uat}) in the second round of fitting
these data. Although the corresponding fit results in Tab.~\ref{tab8} are not
very accurate and stable, the best estimate of this kind, i.~e.,
$\omega = 1.61(44)$ at $L_{\mathrm{min}}=64$,
appears to be consistent with the other our best estimates
at the assumption of pure power-law scaling.

\begin{table}
\caption{The MC estimates of $\omega$, obtained by fitting the $\partial U/\partial \beta$
data in Tab.~\ref{tab1} to the ansatz $-(\partial U/\partial \beta)/L = A + B L^{-\omega}$ within
$L \in [L_{\mathrm{min}},2048]$. The quality of theses fits is controlled
by  the $\chi^2/\mathrm{d.o.f.}$ and the goodness $Q$ values in the last two columns.}
\label{tab5}
\begin{center}
\begin{tabular}{|c|c|c|c|}
\hline
\rule[-2mm]{0mm}{7mm}
$L_{\mathrm{min}}$ & $\omega$ & $\chi^2/\mathrm{d.o.f.}$ & $Q$ \\
\hline
24  &  1.1747(95)  &  5.96  &  $8.3 \cdot 10^{-10}$ \\
32  &  1.246(15)   &  2.41  &  0.0074 \\
48  &  1.318(29)   &  1.76  &  0.070  \\
64  &  1.393(47)   &  1.44  &  0.175  \\
96  &  1.512(96)   &  1.35  &  0.220  \\
128  &  1.81(19)   &  0.77  &  0.591  \\
\hline
\end{tabular}
\end{center}
\end{table}

\begin{table}
\caption{The MC estimates of $\omega$, obtained by fitting the $\partial U/\partial \beta$
data in Tab.~\ref{tab1} to the ansatz
$-(\partial U/\partial \beta)/L = A + B L^{-\omega}+C L^{-2}$ within
$L \in [L_{\mathrm{min}},2048]$. The quality of theses fits is controlled
by  the $\chi^2/\mathrm{d.o.f.}$ and the goodness $Q$ values in the last two columns.}
\label{tab6}
\begin{center}
\begin{tabular}{|c|c|c|c|}
\hline
\rule[-2mm]{0mm}{7mm}
$L_{\mathrm{min}}$ & $\omega$ & $\chi^2/\mathrm{d.o.f.}$ & $Q$ \\
\hline
8  &  1.2965(84)  &  5.85  &  $6.1 \cdot 10^{-11}$ \\
12  &  1.398(16)   &  1.70  &  0.060 \\
16  &  1.434(28)   &  1.76  &  0.084  \\
24  &  1.567(58)   &  1.07  &  0.383  \\
32  &  1.59(10)   &  1.18  &  0.304  \\
\hline
\end{tabular}
\end{center}
\end{table}

\begin{table}
\caption{The MC estimates of $\omega$, obtained by fitting the $\partial U/\partial \beta$
data in Tab.~\ref{tab2} to the ansatz $-(\partial U/\partial \beta)/L = A + B L^{-\omega}$ within
$L \in [L_{\mathrm{min}},2048]$. The quality of theses fits is controlled
by  the $\chi^2/\mathrm{d.o.f.}$ and the goodness $Q$ values in the last two columns.}
\label{tab7}
\begin{center}
\begin{tabular}{|c|c|c|c|}
\hline
\rule[-2mm]{0mm}{7mm}
$L_{\mathrm{min}}$ & $\omega$ & $\chi^2/\mathrm{d.o.f.}$ & $Q$ \\
\hline
16  &  0.4472(87)  &  7.52  &  $4.4 \cdot 10^{-14}$ \\
24  &  0.497(12)   &  5.27  &  $2.2 \cdot 10^{-8}$ \\
32  &  0.553(17)   &  3.17  &  0.00045 \\
48  &  0.629(26)   &  1.61  &  0.105  \\
64  &  0.621(33)   &  1.80  &  0.072  \\
96  &  0.715(54)   &  1.32  &  0.234  \\
128  &  0.850(85)   &  0.69  &  0.657  \\
\hline
\end{tabular}
\end{center}
\end{table}

\begin{table}
\caption{The MC estimates of $\omega$, obtained by fitting the $\partial U/\partial \beta$
data in Tab.~\ref{tab2} to the ansatz
$-(\partial U/\partial \beta)/L = A + B L^{-\omega}+C L^{-1}$ within
$L \in [L_{\mathrm{min}},2048]$. The quality of theses fits is controlled
by  the $\chi^2/\mathrm{d.o.f.}$ and the goodness $Q$ values in the last two columns.}
\label{tab8}
\begin{center}
\begin{tabular}{|c|c|c|c|}
\hline
\rule[-2mm]{0mm}{7mm}
$L_{\mathrm{min}}$ & $\omega$ & $\chi^2/\mathrm{d.o.f.}$ & $Q$ \\
\hline
8  &  0.533(21)  &  6.91  &  $1.5 \cdot 10^{-13}$ \\
12  &  0.729(35)   &  2.55  &  0.0022 \\
16  &  0.819(49)   &  2.11  &  0.017  \\
24  &  1.039(93)   &  1.37  &  0.186  \\
32  &  1.08(14)   &  1.51  &  0.139  \\
48  &  0.92(21)   &  1.61  &  0.117  \\
64  &  1.61(44)   &  1.05  &  0.392  \\
\hline
\end{tabular}
\end{center}
\end{table}

Allowing for the logarithmic corrections,
we have verified the consistency of the data at $U=U^*$
with the $\propto L^{-4/3}/\ln L$
corrections to scaling. We have tested the
quality of the fit in the case if $L^{-\omega}$ is
replaced by $L^{-4/3}/\ln L$
in our best power-law estimation, presented in Tab.~\ref{tab6}.
In this case, we have $\chi^2/\mathrm{d.o.f.}=1.56$ and $Q=0.103$
for the fit within $L \in [24,2048]$. This fit appears to be
somewhat worse than the corresponding fit with $Q=0.383$ in Tab.~\ref{tab4},
but still acceptable. It is interesting to mention that the
fit with the $\propto L^{-7/4}\ln L$ correction (as
one of the options following from the analysis in
Sec.~\ref{sec:chi}) is relatively
good in this case, i.~e., it has
$\chi^2/\mathrm{d.o.f.}=1.03$ and $Q=0.417$.

The fits to~(\ref{eq:Uatlog})
with $\omega=4/3$ and $\kappa=-1$, as well as with
$\omega=7/4$ and $\kappa=1$
are relatively good at $U=2$. Namely, for the fits over
$L \in [64,2048]$,
we have $\chi^2/\mathrm{d.o.f.}=0.92$, $Q=0.496$
in the first case and $\chi^2/\mathrm{d.o.f.}=0.94$,
$Q=0.482$ in the second case, which can be compared with
the corresponding values in Tab.~\ref{tab8}.

\subsection{Estimation from the pseudocritical couplings}

It is also possible to estimate $\omega$ from the
scaling of the pseudocritical couplings
$\widetilde{\beta}_c(L)$. As mentioned in Sec.~\ref{sec:chi}, their scaling is asymptotically consistent with $U=F(tL^{1/\nu})$. Corrections to scaling
have to be included here for the estimation of $\omega$, i.~e.,
\begin{equation}
 U=F\left(\left(t+ \mathcal{O}\left(t^2 \right) \right) L^{1/\nu} \right) + \mathcal{O}\left(L^{-\omega} \right)
\label{eq:uu}
 \;,
\end{equation}
where analytical corrections to scaling are included
via the $\mathcal{O}\left(t^2 \right)$ term, whereas
the $\mathcal{O}\left(L^{-\omega} \right)$ term
represents the non-analytical correction.
The scaling function $F(z)$ is regular at $z=0$,
so that it can be expanded in powers of $z$, i.~e.,
\begin{equation}
 F(z) = F(0) + z \,F'(0)  + \frac{1}{2} z^2 \,
 F''(0) + \cdots \;,
 \label{eq:ff}
\end{equation}
noting that $F(0) = U^*$ holds.
We can calculate $t$ as a function of $L$ from~(\ref{eq:uu}) and~(\ref{eq:ff}) at a fixed $U$.
Recalling that
$\widetilde{\beta}_c(L) - \beta_c \propto t(L)$,
we find from these equations
$\widetilde{\beta}_c(L) - \beta_c \sim L^{-\frac{1}{\nu}-\omega}$ at $U=U^*$, provided that $\frac{1}{\nu} + \omega < 2 \omega$ holds, which is obviously true in our 2D case.
At $U \ne U^*$, there are also two additional terms
$\mathcal{O}\left(L^{-1/\nu} \right)$ and $\mathcal{O}\left(L^{-2/\nu} \right)$,
which reduce to the $\propto L^{-1}$ and
$\propto L^{-2}$ corrections in our 2D case. It makes the
estimation of $\omega$ very challenging. Therefore we focus only on the $U=U^*$ case, using the ansatz
\begin{equation}
 \widetilde{\beta}_c(L) = \beta_c + A\,
 L^{-1-\omega} \, (\ln L)^{\kappa} \,,
 \label{eq:betafit}
\end{equation}
for testing of the pure power-law scenario
with $\kappa=0$ and allowing also for the logarithmic
corrections.

The fitting results over $L \in [L_{\mathrm{min}},2048]$
at $\kappa=0$ are presented in Tab.~\ref{tab9}.
The quality of these fits is acceptable for
$L_{\mathrm{min}} \ge 48$. Formally, the fit with $L_{\mathrm{min}} = 48$
is the best one, yielding $\omega=1.5019(91)$. However, this value is the smallest
one among those ones provided by the set of acceptable fits in Tab.~\ref{tab9}.
Noting that there might be a small systematic shift of the estimated $\omega$
value with increasing $L_{\mathrm{min}}$ (however, masked by statistical errors),
we have chosen $\omega=1.509(14)$, obtained at $L_{\mathrm{min}} = 64$,
as our final best estimate from this data set. It is motivated by the fact that
this fit is almost as good as that one at $L_{\mathrm{min}} = 48$. Still, the error
bars are slightly larger (which increases the chance that the estimated $\omega$
value is correct within the stated error bars).
The value of $\omega$ agrees well with other acceptable values
in Tab.~\ref{tab9}, as well as with all other best
estimates in our paper made at the assumption of the pure power-law scaling.

\begin{table}
\caption{The MC estimates of $\omega$, obtained by fitting the pseudocritical couplings
in Tab.~\ref{tab1} to the ansatz
$\widetilde{\beta}_c(L) = \beta_c + A\,
 L^{-1-\omega}$ within
$L \in [L_{\mathrm{min}},2048]$. The quality of these fits is controlled
by  the $\chi^2/\mathrm{d.o.f.}$ and the goodness $Q$ values in the last two columns.}
\label{tab9}
\begin{center}
\begin{tabular}{|c|c|c|c|}
\hline
\rule[-2mm]{0mm}{7mm}
$L_{\mathrm{min}}$ & $\omega$ & $\chi^2/\mathrm{d.o.f.}$ & $Q$ \\
\hline
32  &  1.4501(48)  &  5.99  &  $3.7 \cdot 10^{-9}$ \\
48  &  1.5019(91)  &  1.28  &  0.243 \\
64  &  1.509(14)  &  1.38  &  0.198  \\
96  &  1.531(29)   &  1.47  &  0.172  \\
128 &  1.511(46)   &  1.67  &  0.124  \\
\hline
\end{tabular}
\end{center}
\end{table}

We have tested also the consistency with $\omega=4/3$.
Just as in the case with the susceptibility data at $U=U^*$,
no consistency is observed at $\kappa=0$ and $\kappa=1$
in~(\ref{eq:betafit}), assuming $\omega=4/3$,
but the data can be reasonably well fit with $\kappa=-1$ in this case. The fit
results are collected in Tab.~\ref{tab10}.
Comparing with the results in Tab.~\ref{tab9},
we can see that the quality of the fits in Tab.~\ref{tab10} is significantly
lower for $L_{\mathrm{min}}=48$ and $L_{\mathrm{min}}=64$. However, it is even higher
for $L_{\mathrm{min}}=96$ and $L_{\mathrm{min}}=128$.

\begin{table}
\caption{The $\chi^2/\mathrm{d.o.f.}$ and $Q$ values for the
fits of the pseudocritical couplings in Tab.~\ref{tab1} to the ansatz
$\widetilde{\beta}_c(L) = \beta_c + A\,
 L^{-7/3}/\ln L$ within
$L \in [L_{\mathrm{min}},2048]$.}
\label{tab10}
\begin{center}
\begin{tabular}{|c|c|c|}
\hline
\rule[-2mm]{0mm}{7mm}
$L_{\mathrm{min}}$ & $\chi^2/\mathrm{d.o.f.}$ & $Q$ \\
\hline
48  &  5.98  &  $4 \cdot 10^{-9}$ \\
64  &  2.28  &  0.015  \\
96  &  1.23  &  0.279  \\
128 &  1.40  &  0.200  \\
\hline
\end{tabular}
\end{center}
\end{table}

\begin{figure}
\begin{center}
\includegraphics[width=0.477\textwidth]{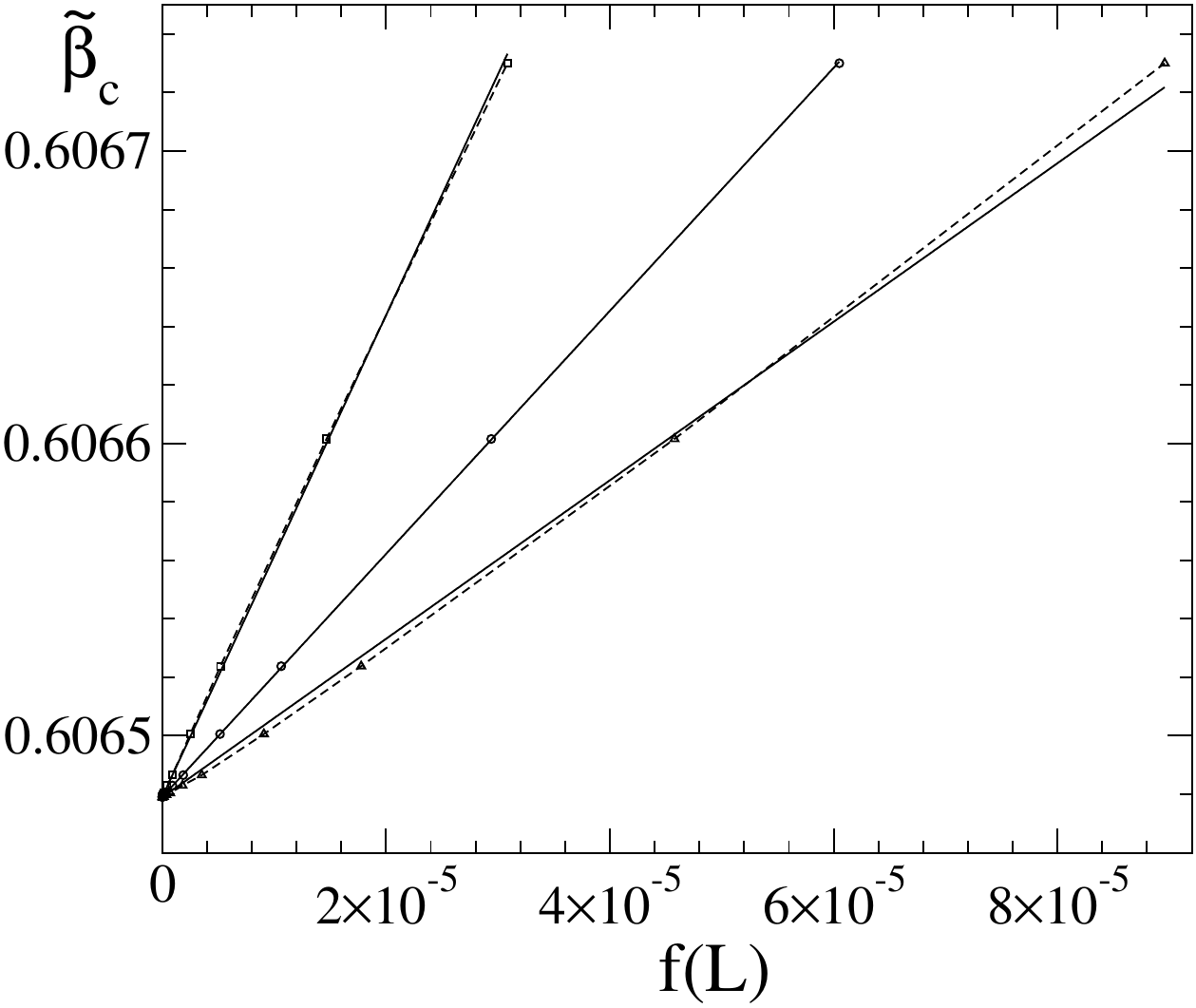}
\hfill
\includegraphics[width=0.49\textwidth]{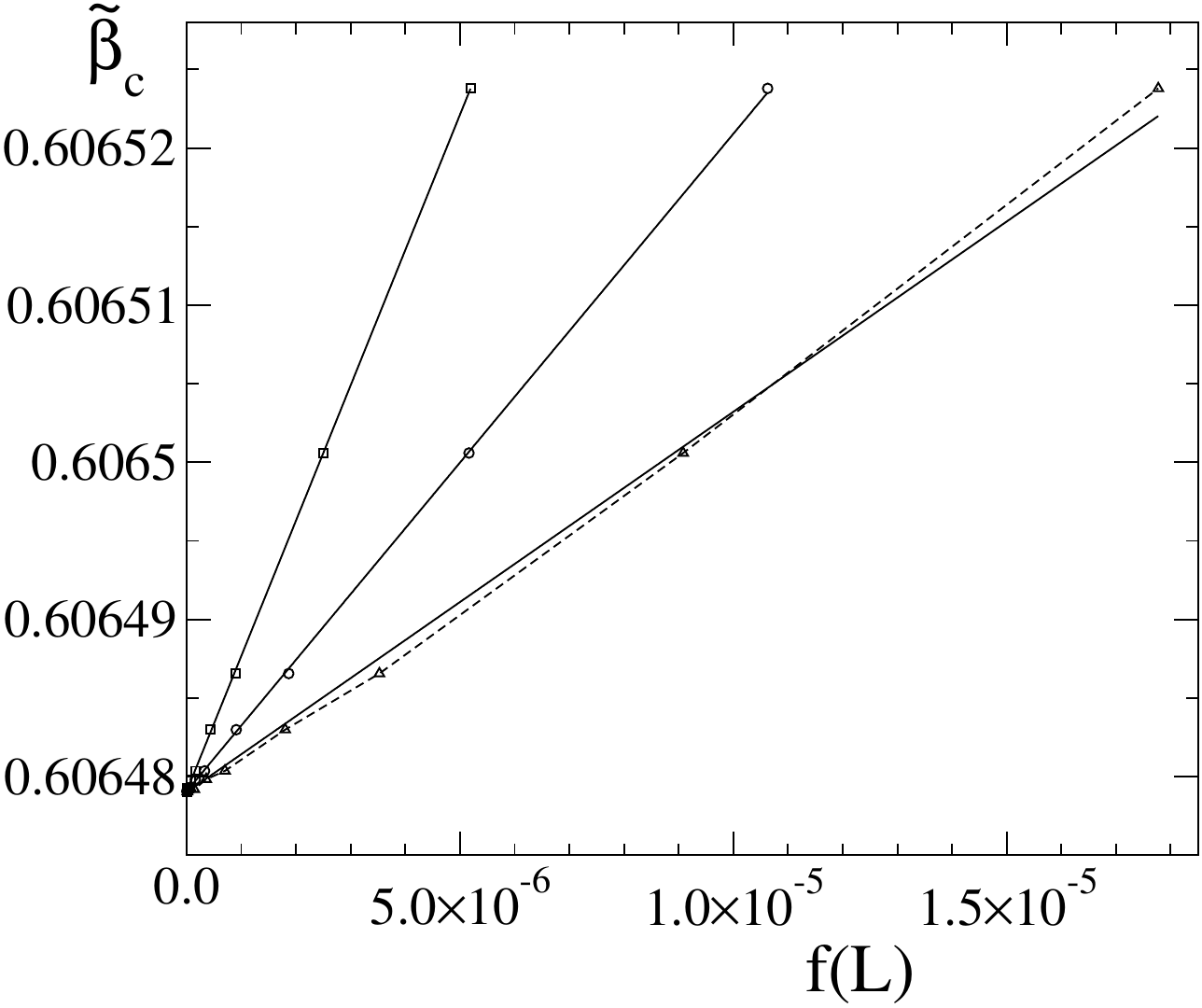}
\end{center}
\caption{The $\widetilde{\beta}_c$ vs $f(L)$ plots
at $U=U^*$ within $L \in [48,2048]$ (left) and $L \in [96,2048]$ (right)
for $f(L)=L^{-2.509}$ (circles), $f(L)=L^{-7/3}/\ln L$
(squares) and $f(L)=0.75 L^{-7/3}$ (triangles).
The statistical errors of one $\sigma$ are smaller than the symbol size
in the left picture. These are about the symbol size or smaller in the right picture.
The linear fits are shown by solid straight lines.}
\label{fig3}
\end{figure}

We have shown and compared the pure power-law fits
with $\omega=1.509$ and $\omega=4/3$, as well as the
fit with $\omega=4/3$ and $\kappa=-1$
over $L \in [48,2048]$ and
$L \in [96,2048]$ in Fig.~\ref{fig3}. In the case of $L \in [48,2048]$,
a slight curvature of the $\widetilde{\beta}_c(L)$ vs
$L^{-7/3}/\ln L$ data curve (dashed line) can be seen,
comparing with the corresponding liner fit. It is in favor
of the pure power-law dependence with $\omega=1.509$.
However, the linear fit to~(\ref{eq:betafit})
with $\omega=4/3$ and $\kappa=-1$ looks perfect
within $L \in [96,2048]$ and even better than the
power-law fit with $\omega=1.509$. Thus, (\ref{eq:betafit})
with $\omega=4/3$ and $\kappa=-1$ can be
the correct asymptotic ansatz. At the same time, the pure
power-law fits with $\omega=4/3$ are not good, as the
$\widetilde{\beta}_c$ vs $0.75 L^{-7/3}$ plots show a remarkable
curvature. These fits are not acceptable according to the formal
criteria: $\chi^2/\mathrm{d.o.f}=39.65$,
$Q=5.3 \cdot 10^{-79}$ for $L \in [48,2048]$
and $\chi^2/\mathrm{d.o.f}=7.88$,
$Q=1.2 \cdot 10^{-10}$ for $L \in [96,2048]$.

Finally, we would like to mention that the fits
with $\omega=7/4$ and $\kappa=1$ are acceptable.
In this case, we have $\chi^2/\mathrm{d.o.f.}=1.51$, $Q=0.128$
and $\chi^2/\mathrm{d.o.f.}=1.42$, $Q=0.180$ for
the fits over $L \in [48,2048]$ and $L \in [96,2048]$, respectively.
\\

\subsection{Summary of MC estimations and comparison}
\label{sec:MCcom}

In summary of our MC estimation, we have found a set of estimates for the
correction-to-scaling exponent $\omega$, which are valid if the corrections
to scaling have a pure power-law form. The most accurate of them are
$\omega=1.509(14)$ and $\omega=1.546(30)$ obtained from our MC data of
the pseudocritical couplings and susceptibility, respectively,
for $U=U^*$ listed in Tab.~\ref{tab1}. Three other estimates of this set
are $\omega = 1.540(69)$, obtained from the susceptibility data at $U=2$ (Tab.~\ref{tab2}), as well
$\omega=1.567(58)$ and $\omega=1.61(44)$,
obtained from the $\partial U/\partial \beta$ data
at $U=U^*$ and $U=2$, respectively. All these values are consistent within the error bars.
These estimates agree well with the result $\omega=1.6(2)$
of the $\epsilon$-expansion reported in~\cite{GZJ}, but are less well consistent
with the more recent result of this expansion $\omega=1.71(10)$ in~\cite{Shalaby}.

Allowing for logarithmic corrections, our data appear to be consistent with
$\omega=4/3$, if the leading non-analytical correction to scaling is modified by the
logarithmic factor of the form $1/\ln L$, and with $\omega=7/4$, if
this logarithmic factor is just $\ln L$. In the first case, the exponent $\omega$
is such one, which can be expected from the results of~\cite{AA}, noting that some
logarithmic correction could be, indeed, expected in this case -- see Sec.~\ref{intro}. In the second case, the value of $\omega$ agrees well
with $1.71(10)$ obtained from the $\epsilon$-expansion in~\cite{Shalaby}. There is only a question about the existence
of the logarithmic correction in this case.

The refined here
estimations do not confirm the existence of corrections to scaling with
the full set of exponents $n/4$, discussed in~\cite{our2D}, where $n$ is an integer
number, although it confirms the existence of non-integer exponents suggested
earlier~\cite{our2D}.
Note that only integer correction exponents
are usually expected in the
2D Ising model,
where $\omega=2$~\cite{CFTrecent}. It corresponds to the limit
$\lambda \to \infty$. In our previous study~\cite{our2D}, we have
considered $\lambda = 0.1, 1, 10 $, noting that the $\lambda=0.1$ case
gives a better chance to identify corrections with non-integer
$\omega$. Here we have focused just on this case. There might be a danger
that the convergence to the correct asymptotic exponents becomes slower with
decreasing $\lambda$, since $\lambda=0$ corresponds to the free-field or Gaussian
model. However, the normalized quantities $\chi/L^{7/4}$ and $(\partial U/\partial \beta)/L$
show the expected convergence to constant values, showing that the asymptotic scaling regime
is reached in our simulations at $\lambda=0.1$. In fact, $\omega=2$ is expected
from the conformal field theory (CFT)~\cite{CFTrecent} in the 2D $\varphi^4$ model
just as in the 2D Ising model, so that the non-integer values of $\omega$
provided by our fits
require further examination from the point of view of CFT. Perhaps, one should allow
that CFT does not entirely describe all corrections to scaling.
Such a possibility has been discussed in~\cite{ourNPRG}.

Our actual MC results serve as a basis for a critical
reconsideration of some earlier theoretical conjectures and scaling assumptions
in~\cite{K_Ann01,our2D}. It is done in the following two sections.

\section{The singularity of specific heat and
the two-point \\ correlation function}
\label{sec:CV}

In~\cite{our2D}, theoretical arguments have been provided concerning corrections
to scaling in the two-point correlation function, based on the continuous
version of the $\varphi^4$ model with

\begin{equation} \label{eq:H}
\frac{\cal{H}}{k_{\mathrm{B}} T}= \int \left( r_0 \varphi^2(\bm{x}) + c (\nabla \varphi(\bm{x}))^2
+ u \varphi^4(\bm{x}) \right) {\rm d} \bm{x} \;,
\end{equation}
where the order parameter $\varphi(\bm{x})$ is an
$N$--component vector with components $\varphi_i(\bm{x})$ depending on the
coordinate $\bm{x}$.
It is assumed that there exists the upper cut-off parameter $\Lambda$
for the Fourier components of the order-parameter field $\varphi_i(\bm{x})$.
It is assumed that $r_0$ is linear function of $T$, $c$ and $u$ being constants.
In this case, the singular part of specific heat $C_V$ can be written as~\cite{our2D}
\begin{equation}
C_V^{\mathrm{sing}} \propto \xi^{1/\nu} \, \left( \int_{k<\Lambda} [G(\bm{k})- G^*(\bm{k})] {\rm d} \bm{k} \right)^{\mathrm{sing}} \;,
\label{eq:CVsing}
\end{equation}
where $k=\mid \bm{k} \mid$, $\xi$ is the correlation length, and
$G(\bm{k})$ is the Fourier-transformed two-point correlation function having
the value $G^*(\bm{k})$ at the critical point. The scaling hypothesis in the form of
\begin{equation}
G(\bm{k}) = \sum\limits_{i \ge 0} \xi^{(\gamma - \theta_i)/\nu} g_i(k \xi) \;,
\label{eq:sc1}
\end{equation}
has been applied for the calculation of~(\ref{eq:CVsing}), keeping a finite number of terms,
where $g_i(k \xi)$ are continuous scaling functions, which are finite for $0 \le k \xi < \infty$. Here $\theta_0=0$ holds and the term with $i=0$
describes the leading singularity, whereas the terms with $i \ge 1$ represent other
contributions with
correction exponents $\theta_i>0$.
The critical correlation function
\begin{equation}
G^*(\bm{k}) = \sum\limits_{i \ge 0} b_i k^{(-\gamma + \theta_i)/\nu}
\label{eq:sc2}
\end{equation}
is obtained at $\xi \to \infty$, so that there exists a finite limit
\begin{equation}
\lim\limits_{z \to \infty} z^{(\gamma - \theta_i)/\nu} g_i(z) = b_i \;,
\label{eq:sc3}
\end{equation}
where $b_i$ are constant coefficients.

An additional assumption has been made, according to which the correct leading
singularity of $C_V$ with certain constant amplitude is provided by the integration
over the region
$k<\Lambda'$ in~(\ref{eq:CVsing}) in the limit
$\lim\limits_{\Lambda'\to 0} \lim\limits_{\xi\to\infty}$. It implies that this
small-$k$ contribution is
asymptotically (at $\xi \to \infty$) independent of $\Lambda'$ for
small enough values of $\Lambda'$. In other words, we can separate a certain
$\Lambda'$-independent long-wavelength contribution.

A theorem has been proven in~\cite{our2D}, according to which the known logarithmic
singularity of $C_V$ at $d=2$ is possible only if there exists a correction with
the exponent $\omega_i=\theta_i/\nu = 3/4$ at the assumptions made.
This exponent, however, is not confirmed by our current MC estimation.
It rises a question: what could be wrong with these assumptions or
conditions of the theorem? Some numerical tests have been performed in~\cite{our2D}
to verify these assumptions. In particular, it has been checked that~(\ref{eq:CVsing})
really provides the logarithmic singularity of the form $C_V=C \ln \xi$,
where $C$ is a constant.
Moreover, it has been checked that
the logarithmic
singularity is provided separately by the contributions of $0<k<\Lambda'$
and $\Lambda'<k<\Lambda$ for certain finite values of $\Lambda'$.
According to the assumptions made, these contributions, denoted as
$A(\Lambda') \ln \xi$ and $B(\Lambda') \ln \xi$ with
$A(\Lambda')+B(\Lambda') = C$, should be almost independent of $\Lambda'$
for small enough values of $\Lambda'$. The dependence of $A(\Lambda')$ and
$B(\Lambda')$ has not been really tested for small $\Lambda'$ values,
relying on the intuitive assumption that $A(\Lambda') \ln \xi$
has to be asymptotically $\Lambda'$-independent
at small $\Lambda'$ and $\xi \Lambda' \to \infty$ as a long-wavelength contribution,
where only vanishingly small wave vectors are relevant at $\xi \Lambda' \to \infty$.

As we currently do not see a reason why the scaling relations
(\ref{eq:sc1})--(\ref{eq:sc3}) could be not valid for the estimation of
the leading long-wavelength contribution, we propose that the reason of the
discrepancy with the MC estimation of $\omega$ is that $A(\Lambda')$ essentially
depends on $\Lambda'$ for small $\Lambda'$ contrary to the assumption made in~\cite{our2D}.
From the formal point of view, it is even very likely that $-B(\Lambda')$ and
$A(\Lambda')$ diverge at $\Lambda' \to 0$ according to the scaling behavior
shown in Fig.~6 in~\cite{our2D}, where $\mid k \, \partial G(\bm{k})/\partial \beta \mid$
tends to increase rapidly for small $k$ in vicinity of the critical point (at $t=0.005$).
This allows also to relate the leading logarithmic singularity of $C_V$ to
the leading singularity of $G(\bm{k})$ rather than to correction terms,
which is quite plausible from an intuitive point of view.

Keeping only the leading term in~(\ref{eq:sc2}) and~(\ref{eq:sc3}), we have
\begin{equation}
 G(\bm{k}) - G^*(\bm{k}) \approx \xi^{\lambda} \tilde{g}(k \xi) \,,
\label{eq:starp}
\end{equation}
where $\lambda = \gamma/\nu = 2 - \eta$, $\eta$ being the critical exponent
describing the behavior of the critical correlation function,
i.~e., $G^*(\bm{k}) \approx b_0 k^{-\lambda}=b_0 k^{-2+\eta}$ at $k \to 0$, and
\begin{equation}
 \tilde{g}(k \xi) = g(k \xi) -b_0 (k \xi)^{-\lambda} \,.
\end{equation}
Inserting~(\ref{eq:starp}) into~(\ref{eq:CVsing}), where we formally set
$\Lambda \to \Lambda'$ for the estimation of the long-wavelength contribution
$C_V^{\mathrm{sing}}(\Lambda')$, we obtain
\begin{equation}
 C_V^{\mathrm{sing}}(\Lambda') \propto \xi^{\frac{1}{\nu}+\lambda}
 \int\limits_0^{\Lambda'} \tilde{g}(k \xi) k^{d-1} {\rm d} k =
 \xi^{\frac{1}{\nu}+\lambda-d}
 \int\limits_0^{\xi \Lambda'} \tilde{g}(z) z^{d-1} {\rm d} z =
 \xi^{\frac{1}{\nu}+\lambda-d} \, F(\xi \Lambda') \,,
 \label{eq:CV}
\end{equation}
where $F(\xi \Lambda')$ is some function of the given argument.

According to the arguments in~\cite{our2D}, $F(\xi \Lambda')$ should be independent
of $\Lambda'$ in the limit
$\lim\limits_{\Lambda'\to 0} \lim\limits_{\xi\to\infty}$, if~(\ref{eq:CV}) represents the
leading singularity of $C_V$.
It is possible, e.~g., if $F(z)$ tends
to some nonzero constant at $z \to \infty$. However, then
we obtain $C_V^{\mathrm{sing}} \propto \xi^{3/4}$ in two
dimensions instead of the expected logarithmic singularity.
No $\Lambda'$-independent logarithmic singularity can be obtained
from~(\ref{eq:CV}) at $d=2$.
Therefore, the formal conclusion from the considerations in~\cite{our2D} was
that $F(\xi \Lambda')$ vanishes at $\xi \Lambda' \to \infty$ in such a way
that $\xi^{\frac{1}{\nu}+\lambda-d} \, F(\xi \Lambda')$ does not represent
the leading singularity of $C_V$, the latter being provided by a correction-to-scaling
term.

Here we consider another possibility, allowing
that~(\ref{eq:CV}) represents
the leading singularity of $C_V$ with $\Lambda'$-dependent amplitude.
This approximation is applicable at $\xi \Lambda' \to \infty$
for any given $\Lambda'$, which should
be small enough to ensure the validity of~(\ref{eq:CV}).
We have $\frac{1}{\nu}+\lambda-d = 3/4$ at $d=2$, so that the expected
logarithmic singularity is recovered if $F(z) \propto z^{-3/4} \ln z$
holds at $z \to \infty$. It implies that $C_V^{\mathrm{sing}}(\Lambda')$ is proportional
to $(\Lambda')^{-3/4}$ at $\xi \to \infty$. To show that this new possibility
is physically meaningful, we point to the fact that
$\xi^{\frac{1}{\nu}+\lambda-d} \, F(\xi \Lambda')$
can be seen as a long-wavelength
contribution according to some criterion.

The old criterion
in~\cite{our2D} states
that the  long-wavelength contribution is such one, which is
independent of $\Lambda'$. The actual new criterion states that a long-wavelength
contribution is such one, for which a small-$k$ subregion
$0 < k < {\cal C} / \xi$ is relevant at $\xi \to \infty$ for any positive
constant ${\cal C}$.
In other words, this contribution  has to be changed significantly,
if this subregion is cut off. Such a cutting procedure
essentially modifies the result~(\ref{eq:CV}) from
$\xi^{\frac{1}{\nu}+\lambda-d} F(\xi \Lambda')$  to
$\xi^{\frac{1}{\nu}+\lambda-d} (f({\cal C}) + F(\xi \Lambda'))$, where
$f({\cal C})$ is a function of ${\cal C}$.
Hence, (\ref{eq:CV}) can be seen as a long-wavelength contribution according to
the new criterion proposed.

The new criterion can be seen as a generalization
of the old one. Indeed, if we deal with contributions,
which come just from the region $k \sim 1/\xi$ and are independent of $\Lambda'$,
then these contributions are modified by the cutting of a subregion
of $k \sim 1/\xi$. Hence, theses are the long-wavelength contributions
according to both criteria. A relevant example is
$C_V^{\mathrm{sing}}(\Lambda')$ above the upper critical dimension, i.~e., at $d>4$.
In this case, the correct leading singularity
$C_V^{\mathrm{sing}} \propto \xi^{4-d} \propto \, \mid t \mid^{(d-4)/2}$ is recovered at
a constant (nonzero) asymptotic value of $F(z)$ at $z \to \infty$.
The quantity $C_V^{\mathrm{sing}}(\Lambda')$ appears to be $\Lambda'$-independent in this case.

Summarising the discussion of this section, we have proposed here a new scaling conception
for the calculation of the leading singularity of $C_V$ in the actually considered
continuous $\varphi^4$ model. It is based on a more flexible scaling assumption
concerning the behavior of the function $F(\xi \Lambda')$ in~(\ref{eq:CV}) as compared
to the earlier consideration in~\cite{our2D}. Namely, here we allow that $F(z)$ essentially
depends on $z$ at $z \to \infty$, showing that~(\ref{eq:CV}) still can be seen
as a physically meaningful long-wavelength contribution in this case.
It shows how the correct leading singularity of $C_V$ can
be obtained without any conditions for the correction-to-scaling exponents.
Hence, no correction exponent $3/4$ is expected from the actual consideration at $d=2$.
It resolves the apparent contradiction between the theory and the MC estimation
due to the earlier assumptions in~\cite{our2D}.

\section{A renewed analysis by grouping of Feynman diagrams}
\label{sec:GFD}

A method of grouping Feynman diagrams (GFD) has been proposed in~\cite{K_Ann01} for the
analysis of critical exponents in the $\varphi^4$ model. Later on~\cite{PTP},
this method has been extended to the analysis of the Goldstone mode singularity.
The analysis in these papers refers basically to the case $d <4$,
and all considerations in this section also refer to this case.

The analysis in~\cite{K_Ann01} requires a significant reconsideration from the
point of view of the actual MC results in Sec.~\ref{sec:MCes} and findings in Sec.~\ref{sec:CV}. The grouping of Feynman diagrams in~\cite{K_Ann01}
has been done in a formally rigorous way. Only the
analysis of the resulting equations is somewhat problematic.

The analysis at $T \to T_c$, leading to a certain
prediction for the possible values of the critical
exponents $\gamma$ and $\nu$~\cite{K_Ann01}, is largely based
on the old rather than the new scaling assumption concerning the
relevant long-wavelength contribution to specific heat $C_V$, discussed
in Sec.~\ref{sec:CV}. The new scaling conception is more flexible and,
therefore, it does not lead to restrictions for the possible
values of the critical exponents. Hence, the prediction for $\gamma$
and $\nu$ in~\cite{K_Ann01}
is not further supported by the renewed analysis.

The method of analysis of the two-point correlation
function $G(\bm{k})$ at the critical point in~\cite{K_Ann01} could be meaningful,
as regards the leading asymptotic behavior, at least.
The equation for $G(\bm{k})$ at $T=T_c$ has the form
\begin{equation}
 \frac{1}{2G(\bm{k})} = R(\bm{k}) - R(\bm{0}) + ck^2 \,,
\label{eq:G}
 \end{equation}
where $R(\bm{k})$ is represented by a resummed sum of infinitely many the
so-called skeleton diagrams in~\cite{K_Ann01}. In the formal analysis
of~\cite{K_Ann01}, Eq.~(\ref{eq:G}) contains also a nonperturbative
correction term, which is neglected in the limit of small $u$ and
small $k$ considered throughout that paper. The quantity $R(\bm{k})$ is evaluated
at $G(\bm{k})$ in the form of~(\ref{eq:sc2}), resulting in
\begin{equation}
 R(\bm{k}) - R(\bm{0}) = \sum\limits_{i \ge 0} k^{\lambda+\omega_i}
  \, \Psi_i(\Lambda/k) \,,
\label{eq:R}
\end{equation}
where $\omega_i = \theta_i/\nu$ and $\Psi_i(\Lambda/k)$ are scaling functions,
which depend on $\Lambda/k$, as well as on other arguments
($d$, $N$, $\lambda$, $\omega_i$ and $u$), which are skipped here for brevity.
According to the assumption in~\cite{K_Ann01}, these scaling functions
can be approximated by $\Lambda$-independent constants at small $k$.
It is justified if the relevant
long-wavelength contributions come from the region $q_{\ell} \sim k$,
where $\bm{q}_{\ell}$ denotes the $\ell$-th internal wave vector contained in the
diagrams of $R(\bm{k})$.

If only the leading term is taken into account,
then~(\ref{eq:R}) contains only one scaling function $\Psi_0(\Lambda/k)$.
It is expected that $\Psi_0(\Lambda/k)$
tends to a certain finite positive value at $\Lambda/k \to \infty$
to ensure that~(\ref{eq:G}) has a positively defined solution in the form of
$G(\bm{k}) \propto k^{-\lambda}$ at $k \to 0$ with $\lambda <2$.
However, the applicability of this analysis cannot be rigorously proven,
as we cannot really sum up all the diagrams and aware that
$\Psi_0(\Lambda/k)$ does indeed tend to a finite positive value.
Alternatively, one can formally set $\Lambda = {\cal C} k$ or
$\Lambda/k = {\cal C}$, where ${\cal C}$ is a finite positive
constant, considering this as some approximation.

If corrections to scaling are taken into account, then
at least one of the terms $R(\bm{k}) - R(\bm{0})$
and $1/(2G(\bm{k}))$ contains a correction
with the exponent $\omega_1=2-\lambda=\eta$ to compensate
the term $ck^2$ in~(\ref{eq:G}).
At the assumption that $\Psi_i(\Lambda/k)$ can be considered as
being independent of $\Lambda/k$,
such a correction term in $R(\bm{k}) - R(\bm{0})$
can arise only from the corresponding correction term in $G(\bm{k})$,
in accordance with the scaling analysis of the diagrams in~\cite{K_Ann01}.
Hence, this assumption
leads to the conclusion that
$G(\bm{k})$ must contain a correction with the exponent $\omega_1=\eta$
to satisfy~(\ref{eq:G}).
Such a correction exponent, i.~e., $\omega_1=1/4$ in two dimensions,
is not confirmed by our actual MC estimation in Sec.~\ref{sec:MCes}.
It already indicates that the considered here assumption is not valid
for the correction terms. The formal reason for this discrepancy could be the fact
that other contributions, apart from those of the region  $q_{\ell} \sim k$,
are important.

This discrepancy can be easily resolved,
including the analytical background contribution into the consideration.
It shows up as an additive $\sim k^2$ analytical term in $R(\bm{k}) - R(\bm{0})$,
which comes from the whole integration region  $q_{\ell} \le \Lambda$.
As we have tested, considering only the simplest skeleton
diagrams, this additive contribution can compensate the term $ck^2$ in~(\ref{eq:G}).
In this case, no correction with the exponent $\omega_1=\eta$ appears
in $G(\bm{k})$.

An approximate qualitative analysis can be performed, including the
analytical background contribution in addition
to the contributions of the region $q_{\ell} \sim k$.
The latter contributions can be evaluated by formally setting
$\Lambda = {\cal C} k$, where ${\cal C}$ is a positive constant. It
reduces to the formal setting $\Psi_i(\Lambda/k) \to \Psi_i({\cal C})$
in~(\ref{eq:R}). The analytical background contribution
shows up as an additional term of the form
$\tilde{c} k^2 + o \left( k^2 \right)$, where $\tilde{c}$ is a constant. The expected
compensation of the term $ck^2$ in~(\ref{eq:G}) takes place at $\tilde{c}=-c$.

The amplitude $a$ in $G(\bm{k}) = a k^{-\lambda}$ at $k \to 0$
is not uniquely defined from the asymptotic solution of~(\ref{eq:G}),
if only the leading terms are taken into account. Namely,
setting $G(\bm{k}) = a k^{-\lambda}$, we have
$(2a)^{-1} k^{\lambda}$ in the left hand side of~(\ref{eq:G})
and $(2a)^{-1} k^{\lambda} \, \widetilde{\Psi}_0(\Lambda/k,\lambda) =
(2a)^{-1}k^{\lambda} \, \widetilde{\Psi}_0({\cal C,\lambda})$ in the right hand side
of this equation, where $\widetilde{\Psi}_0 = 2a \Psi_0$ is the
normalized scaling function, which is independent of $a$.
It follows from the diagrammatic analysis in~\cite{K_Ann01}
with inclusion of only the leading terms.
Hence, the exponent $\lambda$ is obtained from
the condition $\widetilde{\Psi}_0({\cal C,\lambda})=1$,
whereas the amplitude $a$ remains uncertain.

An approximate estimation of the coefficient $\tilde{c}$ is possible for the
simplest skeleton diagrams (including only the first expansion
term in Eq.~(29) of~\cite{K_Ann01}), at least. It turns out that
this coefficient depends on the amplitude $a$ in such an estimation,
i.~e., we have $\tilde{c} = \tilde{c}(a)$.
Since the amplitude $a$ is not uniquely determined yet,
we can use it as a free parameter to find a solution
satisfying the expected relation $\tilde{c} = -c$ at a certain value of $a$.
The absence of the correction with the
exponent $\omega_1=\eta$ in $G(\bm{k})$ implies that this is the
physically meaningful solution. In this case,
 $\omega = 2 \lambda - d = 4-d-2 \eta$
shows up as the leading correction-to-scaling exponent in our approximation.
It follows from the scaling analysis of the diagrams in~\cite{K_Ann01},
where two independent correction exponents, i.~e.,
$\eta$ and $2 \lambda -d$, have been found.
Only the second one survives at the condition $\tilde{c} = -c$.

In analogy with the analysis in Sec.~\ref{sec:CV}, the problem can appear
to be unexpectedly complex, and the dependence of the scaling functions on $\Lambda/k$
may also influence the results. Therefore, we consider
the obtained relation between $\eta$ and $\omega$
as an approximation, i.~e.,
\begin{equation}
 \omega \approx 4-d-2 \eta  \qquad  \textrm{for} \quad d < 4 \,,
 \label{eq:relation}
\end{equation}
resulting from an approximate evaluation of $R(\bm{k})-R(\bm{0})$
at a finite value of ${\cal C}$.

Although our approximation scheme, eventually, does not allow to approach
the exact solution of~(\ref{eq:G}) due to the problems discussed,
(\ref{eq:relation}) turns out to be a valid estimate.
Indeed, it gives asymptotically correct result
at $\varepsilon=4-d \to 0$, as well as at $N \to \infty$,
where $\omega \approx \varepsilon$. It gives $\omega \approx 0.928$
for the 3D Ising model (where $\eta \approx 0.036$), which is
comparable with numerical estimates, usually providing $\omega$
around $0.8$, and probably is not worse than the
local potential approximation, giving $\omega \approx 0.65$
(see, e.~g., references in~\cite{ourNPRG}).
The approximation~(\ref{eq:relation}) gives $\omega \approx 1.5$
in two dimensions in a surprisingly good agreement with
our actual best MC estimate $\omega = 1.509(14)$,
 which is valid if the leading non-analytical
 correction to scaling has the pure power-law form.
 So, an accurate agreement is, indeed, surprising since~(\ref{eq:relation})
 is just an approximation, which is not very accurate in the three dimensions.
 Allowing for logarithmic corrections,
 the correct $\omega$ value might be $4/3$, as discussed throughout
 our paper. In this case, the value $1.5$ of $\omega$ at $d=2$
 provided by~(\ref{eq:relation}) appears to be somewhat overestimated,
 just as in the 3D case.

 The discussion above reveals essential challenges in the analysis
 of the diagrammatic equations derived in~\cite{K_Ann01}.
 Nevertheless, this GFD method can be useful in some cases.
 Particularly, it is suited for the analysis of the large $N$ limit,
 where asymptotically correct results are provided by relatively
 simple equations of this approach, containing only the first diagram
 in the expansion~(29) of~\cite{K_Ann01}.
 The detailed analysis of this case goes beyond
 the scope of this paper. Nonetheless, we can mention the fact that the known asymptotic result for $\eta$ at large $N$ and
 $2 < d <4$~\cite{AH}, i.~e.,
\begin{equation}
  \eta = \frac{2}{N} \; \frac{\sin(\pi d/2) \, \Gamma(d-1) \,
  (1-[4/d])}{\pi \left[ \Gamma(d/2)  \right]^2} \,,
\label{eq:eta}
  \end{equation}
is easily recovered from these GFD equations.
Therefore, it is also very interesting to look what these equations
give for the Goldstone mode singularity at large $N$.

\section{Summary and discussion}
\label{sec:summary}

Based on our MC analysis, the following three theoretical possibilities
concerning the asymptotic behavior of the leading non-analytical correction to scaling in the 2D $\varphi^4$ model can be considered:
\begin{itemize}
 \item[(i)] In view our best MC estimates (each from his own data set)
 valid for $\propto L^{-\omega}$ scaling
of the non-analytical correction term, i.~e.,
 $\omega=1.509(14)$, $\omega=1.546(30)$,
 $\omega = 1.540(69)$, $\omega=1.567(58)$ and $\omega=1.61(44)$,
the correction-to-scaling exponent $\omega$ is close to $1.5$
if this scaling has, indeed, a pure power-law form.
The most accurate estimate $\omega=1.509(14)$ among these
ones suggests that $\omega$ could be just $1.5$,
but we do not stress on this value,
allowing for a slightly larger value as very probable.
This concept is plausible from the point of
view that the power-law scaling is the usual one.
The only problem is that such
values of $\omega$ are currently not confirmed by any exact result.
\item[(ii)] Looking for consistency with some exact result,
we have found that $\omega$ is consistent with $4/3$, provided
that the $\propto L^{-\omega}$ scaling is modified by the
logarithmic correction factor $1/\ln L$. The value $4/3$
has been reported in~\cite{AA} as the value of two coinciding
correction-to-scaling exponents $\theta$ and $\theta'$, describing the fractal
geometry of the critical Potts clusters in the case of the
Ising model ($q=2$) in two dimensions. We have considered
the coincidence of these exponents as the reason for
possible logarithmic correction.
This conception seems to be very plausible
from the point of view that it establishes the consistency with the exact
exponent $4/3$ found in~\cite{AA}. However, we still have some
concerns about its validity, since
it is not rigorously clear whether or not the exponents
$\theta$ and $\theta'$ are applicable to other
quantities apart from those
related to
the fractal geometry of the critical Potts clusters.
This is a relevant question
from the point of view of the CFT discussed in~\cite{PVx},
since the percolation problem of~\cite{AA} is a nonunitary
extension of the theory, and no correction exponent
$4/3$ has been found in the unitary $\varphi^4$ theory, suggesting
$\omega=2$.
Nevertheless, we still consider options with
$\omega \ne 2$
because the observed scaling of various quantities, e.~g.,
$\widetilde{\beta}_c(L)$
at $U=U^*$ (accurately described by $\omega=1.509(14)$
at $\kappa=0$ in~(\ref{eq:betafit}) within $L \in [48,2048]$
-- see Tab.~\ref{tab9} and Fig.~\ref{fig3})
is not predicted by this theory
(suggesting $\omega=2$ and $\kappa=0$ in~(\ref{eq:betafit})),
indicating that the theory itself might be incomplete.
Another problem is that a
pure power-law scaling form
has been proposed in~\cite{AA} at $q=2$ (see Eq.~(11) in~\cite{AA}).
The existence of
a logarithmic correction is just a hypothesis considered
in our paper, since no
consistency of our MC data with $\omega=4/3$ can be established
in the absence of any logarithmic correction
-- see Fig.~\ref{fig1} and Fig.~\ref{fig3}.
\item[(iii)] Considering logarithmic corrections,
one more possibility has been revealed. Namely,
the $\propto L^{-7/4} \ln L$ scaling with $\omega=7/4=1.75$
appears to be pretty well consistent with the data.
This concept can be seen as plausible because
the $\omega=1.75$ version is supported by the
findings in~\cite{Shalaby}, where this value
of $\omega$ is considered to be exact.
However, there are some concerns about the validity
of this conception, too. The basic problem is the
absence of any theoretical argument for the existence
of the logarithmic correction in this case, noting that the
data cannot be acceptably fit with $\omega=1.75$ without
such a correction. A positive aspect, however, is that
the logarithmic correction $\ln L$ is more usual
than the $1/\ln L$ correction, and it really
exists in many cases.
A real support for $\omega=1.75$
comes from the $\epsilon$-expansion in~\cite{Shalaby},
yielding $\omega=1.71(10)$,
rather than from the suggested in~\cite{Shalaby} reference
(our Ref.~\cite{PVx}), claiming $\omega=2$.
The accuracy of the $\epsilon$-expansion at $\epsilon=2$,
however, is questionable.
\end{itemize}

The revealed discrepancies with our earlier expectations have been resolved,
reconsidering the scaling assumptions of~\cite{our2D}
concerning the singularity
of specific heat and the two-point correlation function $G(\bm{k})$
in vicinity of the critical point.
These scaling assumptions are replaced with more general
(more flexible) ones in Sec.~\ref{sec:CV}. This renewed scaling
consideration is crucial also for the GFD analysis in~\cite{K_Ann01}.
The corrected GFD analysis is proposed in Sec.~\ref{sec:GFD}.
It does not allow to predict the possible exact values of the critical
exponents, suggested earlier in~\cite{K_Ann01}. However, it still allows
to propose some approximation~(\ref{eq:relation}), relating the
critical exponents $\omega$ and $\eta$. It gives $\omega \approx 1.5$
for the 2D $\varphi^4$ model.
Despite of the challenges in the general GFD analysis,
we see a perspective in its application
to the large-$N$ limit. It allows to recover easily
the known result~(\ref{eq:eta}). Moreover, it can be further applied to the
analysis of the Goldstone mode singularity at large $N$.

\section*{Acknowledgments}

 This work was made possible by the facilities of the
Shared Hierarchical Academic Research Computing Network
(SHARCNET:www.sharcnet.ca).
The authors acknowledge the use of resources provided by the
Latvian Grid Infrastructure  and High Performance Computing centre
of Riga Technical University.
R. M. acknowledges the support from the
NSERC and CRC program.


\begin{thebibliography}{100}


\bibitem{Amit}
D. J. Amit, \textit{Field Theory, the Renormalization Group, and Critical Phenomena},
World Scientific, Singapore, 1984.

\bibitem{Ma}
S. K. Ma, \textit{Modern Theory of Critical Phenomena},
W. A. Benjamin, Inc., New York, 1976.

\bibitem{Justin}
J. Zinn-Justin, \textit{Quantum Field Theory and Critical Phenomena},
Clarendon Press, Oxford, 1996.

\bibitem{Kleinert}
H. Kleinert, V. Schulte-Frohlinde, \textit{Critical Properties of $\phi^4$ Theories},
World Scientific, Singapore, 2001.

\bibitem{PV}
A. Pelissetto, E. Vicari,
Critical phenomena and renormalization-group theory,
Phys. Rep. 368 (2002) 549--727.

\bibitem{K_Ann01}
J. Kaupu\v{z}s,
Critical exponents predicted by grouping of Feynman diagrams in $\varphi^4$ model,
Ann. Phys. (Berlin) 10 (2001) 299--331.

\bibitem{Shalaby} A. M. Shalaby, Critical exponents of the $O(N)$-symmetric $\phi^4$ model from the
$\varepsilon^7$ hypergeometric-Meijer resummation, Eur. Phys. J. C \textbf{81}, 87 (2021).

\bibitem{MHB86} A. Milchev, D. W. Heermann, K. Binder,
Finite-size scaling analysis of the $\phi^4$ field theory on the square lattice,
J. Stat. Phys. \textbf{44}, 749 (1986).

\bibitem{TC90} R. Toral, A. Chakrabarti,
Numerical determination of the phase diagram for the
$\varphi^4$  model in two dimensions, Phys. Rev. B \textbf{42}, 2445 (1990).

\bibitem{MF92} B. Mehling, B. M. Forrest,
Universality in the critical two-dimensional $\varphi^4$--model, Z. Phys. B \textbf{89}, 89 (1992).

\bibitem{our2D} J. Kaupu\v{z}s, R. V. N. Melnik, J. Rim\v{s}\=ans,
Corrections to finite-size scaling in the $\varphi^4$ model on square lattices,
Int. J. Mod. Phys. C 27 (2016) 1650108.

\bibitem{AGKKW17} N. Anand, V. X. Genest, E. Katz, Z. U. Khandker, M. T. Walters,
RG flow from $\phi^4$ theory to the 2D Ising model,
JHEP \textbf{08}, 056 (2017).

\bibitem{AKY21} S. Akiyama, Y. Kuramashi, Y. Yoshimura,
Phase transition of four-dimensional lattice  theory with tensor renormalization group,
Phys. Rev. D \textbf{104}, 034507 (2021).

\bibitem{nacom24} G. Bighin, T. Enss, N. Defenu,
Universal scaling in real dimension, Nature Communications, published online,
doi.org/10.1038/s41467-024-48537-1 (2024).

\bibitem{Onsager} L.~Onsager, Crystal statistics.
I. A two-dimensional model with an order-disorder transition, Phys.~Rev. \textbf{65} (1944) 117.

\bibitem{Baxter} R. J.~Baxter,
\textit{Exactly Solved Models in Statistical Mechanics},
Academic Press, London, 1989.

\bibitem{HasRev} M. Hasenbusch,
Monte Carlo studies of the three-dimensional Ising model in equilibrium,
Int.~J.~Mod.~Phys. C \textbf{12}, 911 (2001).

\bibitem{Ferrenberg} A. M. Ferrenberg, J. Xu, D. P. Landau,
Pushing the limits of Monte Carlo simulations for the three-dimensional Ising model,
Phys. Rev. E \textbf{97}, 043301 (2018).

\bibitem{mansMC3D} J. Kaupu\v{z}s, R. V. N. Melnik,
Corrections to scaling in the 3D Ising model: A comparison between MC and MCRG results,
 Int. J. Mod. Phys. C, 2350079 (2023).

\bibitem{Wipf} A. Wipf, \textit{High-Temperature and Low-Temperature Expansions.}
In: Statistical Approach to Quantum Field Theory. Lecture Notes in Physics,
vol.~992, Springer, Cham (2021).

\bibitem{BC2001} P. Butera, M. Comi, Critical universality and hyperscaling revisited for Ising models of general spin using extended high-temperature series, Phys. Rev. B \textbf{65}, 144431 (2002).

\bibitem{HTCompostrini} M. Compostrini, A. Pelisseto, P. Rossi, E. Vicari,
25th-order high-temperature expansion results for three-dimensional Ising-like systems on the simple-cubic lattice, Phys. Rev. E \textbf{65}, 066127 (2002).

\bibitem{Showk} S. El-Showk, M. F. Paulos, D. Poland, S. Rychkov, D. Simmons-Duffin, A. Vichi,
Solving the 3D Ising Model with the Conformal Bootstrap, J. Stat. Phys. \textbf{157}, 869 (2014).
\bibitem{bootstrap} D. Poland, D. Simmons-Duffin, The conformal bootstrap,
Nature Physics \textbf{12}, 535 (2016)
\bibitem{CFTrecent} M. Reehorst, Rigorous bounds on irrelevant operators in the 3d Ising model, CFTJHEP \textbf{09}, 177 (2022).
\bibitem{CFK22} H. Chen, A. L. Fitzpatrick, D. Karateev,
Bootstrapping 2d $\phi^4$ theory with Hamiltonian truncation data, JHEP \textbf{02}, 146 (2022).
\bibitem{PVx} P. Calabrese, M. Caselle, A. Celi, A. Pelissetto, E. Vicari,
Non-analyticity of the Callan-Symanzik $\beta$-function of two-dimensional O(N) models,
J. Phys. A: Math. Gen. \textbf{33} (2000) 8155-8170.
\bibitem{AA} A. Aharony, J. Asikainen, Fractal dimensions and corrections to scaling for critical Potts clusters,
Fractals, Vol. 11, Supplementary Issue (February 2003) 3–7.


\bibitem{Hasenbusch} M. Hasenbusch, A Monte Carlo study of leading order scaling
corrections of $\phi^4$ theory on a three-dimensional lattice,
J. Phys. A: Math. Gen. \textbf{32}, 4851 (1999).

\bibitem{Sokal} J. Salas, A. D. Sokal, Universal amplitude ratios in the
critical two-dimensional Ising model on a torus, J. Stat. Phys. \textbf{98}, 551 (2000).

\bibitem{Recipes} W. H. Press, B. P. Flannery, S. A. Teukolsky,
W. T. Vetterling, Numerical Recipes -- The Art of Scientific Computing,
Cambridge University Press, Cambridge, 1989.


\bibitem{Perk} W.P. Orrick, B. Nickel, A.J. Guttmann and J.H.H. Perk,
The susceptibility of the square lattice Ising model: New developments,
J. Stat. Phys. \textbf{102}, 795 (2001).

\bibitem{ourNPRG}  J. Kaupu\v{z}s, R.V.N. Melnik, Original and modified non-perturbative
renormalization group equations of the BMW scheme at the arbitrary order of truncation,
Front. Phys. 11:1182056 (2024).

\bibitem{GZJ} J. C. Le Guillou, J. Zinn-Justin,
Accurate critical exponents from the $\varepsilon$-expansion,
J. Physique Lett. \textbf{46} (1985) L-137 -- L-141.

\bibitem{PTP} J. Kaupu\v{z}s, Longitudinal and Transverse Correlation
Functions in the $\varphi^4$ Model
below and near the Critical Point, Progr. Theor. Phys. \textbf{124}, 613 (2010).

\bibitem{AH} R. Abe, S. Hikami, Critical Exponents and Scaling Relations
in $1/n$ Expansion, Progr. Theor. Phys. \textbf{49}, 442 (1973).





\end{thebibliography}
\end{document}